\documentclass[prd,nofootinbib,preprintnumbers]{revtex4}
\usepackage{dcolumn}
\usepackage{graphicx}
\usepackage{amssymb}
\usepackage{lscape}
\usepackage{amsmath}
\usepackage{psfrag}
\usepackage{epsfig}

\begin{document}
\preprint{LMU-ASC 56/10}

\newcommand{\epsi}{\epsilon}
\newcommand{\be}{\begin{equation}}
\newcommand{\ee}{\end{equation}}
\newcommand{\bse}{\begin{subequations}}
\newcommand{\ese}{\end{subequations}}
\newcommand{\tr}{{\mathrm{tr}}}
\newcommand{\arctanh}{\mathrm{arctanh}}
\newcommand{\BD}{BD}
\newcommand{\AGS}{AGS}
\def\L5{\tilde{\Lambda}}
\def\MM{M_{(4)}}
\def\Wi{W_{\infty}}
\def\Xii{\Xi_{\infty}}
\def\Z{\zeta}
\def\Wo{W_0}
\def\Xio{\Xi_0}

\def\bm{\bar{\mu}}
\def\bl{\bar{\lambda}}
\def\bn{\bar{\nu}}

\renewcommand{\d}{{\mathrm{d}}}

\newcommand{\beq}{\begin{equation}}

\newcommand{\eeq}{\end{equation}}
\newcommand{\de}{\partial}

\newcommand{\ud}{\mathrm{d}}
\newcommand{\bea}{\begin{eqnarray}}
\newcommand{\eea}{\end{eqnarray}}
\newcommand{\ba}{\begin{eqnarray}}
\newcommand{\ea}{\end{eqnarray}}
\def\N{{\cal N}}
\def\a{{\cal C}_1}
\def\b{{\cal C}_2}
\def\Mf{M_{(5)}}
\def\Mp{M_p}
\def\inpu{{\left(i+1\right)}}
\def\inmu{{\left(i-1\right)}}
\def\inpk{{(i+k)}}
\def\jn{{(j)}}
\def\kn{{(k)}}
\def\mkn{{(-k)}}
\def\mkmjn{{(-k-j)}}
\def\ipjn{{(i+j)}}
\def\injd{{(i+j(\Delta))}}
\def\x{{\xi}}

\title{The recovery of General Relativity in massive gravity via the Vainshtein mechanism}

\author{E.~Babichev$^{a,b}$, C.~Deffayet$^{c,d}$, R.~Ziour$^c$}
\affiliation{$^a$ ASC, Department f\"ur Physik, Ludwig-Maximilians-Universit\"at M\"unchen,
Theresienstr. 37, D-80333, Munich, Germany}
\affiliation{$^b$ Institute for Nuclear Research, 60th October Anniversary Prospect, 7a, 117312 Moscow, Russia}
\affiliation{$^c$ APC, UMR 7164 (CNRS-Universit\'e Paris 7), 10 rue Alice Domon et L\'eonie Duquet,75205 Paris Cedex 13, France}
\affiliation{$^d$ IAP (Institut d'Astrophysique de Paris), 98bis Boulevard Arago, 75014 Paris, France}

\begin{abstract}

We study in detail static spherically symmetric solutions of non linear Pauli-Fierz theory. We obtain a numerical solution with a constant density source.  This solution shows a recovery of the corresponding solution of General Relativity via the Vainshtein mechanism. This result has already been presented by us in a recent letter, and we give here more detailed information on it as well as on the procedure used to obtain it. We give new analytic insights upon this problem, in particular for what concerns the question of the number of solutions at infinity. We also present a weak field limit which allows to capture all the salient features of the numerical solution, including the Vainshtein crossover and the Yukawa decay.

\end{abstract}

\date{\today}

\maketitle 

\section{Introduction}
Motivated by the wish to find consistent large distance modifications of gravity, theories of ``massive gravity'' have recently attracted some attention (see e.g. \cite{Rubakov:2008nh} for a review).
Starting from the unique consistent quadratic theory for a Lorentz invariant massive spin 2, the Pauli-Fierz theory (henceforth PF) \cite{Fierz:1939ix}, it is easy to build a simple non linear completion at the price of introducing a non dynamical external metric. We will consider here such a non linear Pauli-Fierz (henceforth NLPF) theory  that we will define precisely below.  One of the crucial issues to be answered in models of massive gravity is how to recover metrics sufficiently close to the ones obtained in Einstein's General Relativity (henceforth GR) to pass standard tests of the latter theory, while at the same time having significant deviations from GR at large distances. A first major obstacle in this way is related to the so-called van Dam-Veltman-Zakharov (vDVZ) discontinuity  \cite{vanDam:1970vg}. Indeed, the vDVZ discontinuity states that the quadratic Pauli-Fierz theory does not have linearized GR as a limit when the mass of the graviton is sent to zero. In particular, the quadratic Pauli-Fierz theory leads to physical predictions so different from those of GR, such as those for light bending around the Sun, that one can rule out this theory based on the simplest tests of GR in the solar system. However, soon after the discovery of the vDVZ discontinuity, it was realized that the discontinuity might in fact disappear in nonlinear Pauli-Fierz theories \cite{Vainshtein:1972sx}. This is because a careful examination of static spherically symmetric solutions of those theories by A. Vainshtein showed that the solutions of the linearized nonlinear Pauli-Fierz theories (i.e. those of simple Pauli-Fierz theory) were only valid at distances larger than a distance scale, the Vainshtein radius $R_V$, which goes to infinity when the mass of the graviton is sent to zero. On the other hand Vainshtein showed that there existed a well behaved (as the mass of the graviton is sent to zero) expansion valid at distances smaller than $R_V$, this expansion being defined as an expansion around the Schwarzschild solution of GR. What Vainshtein did not show is the possibility to join together those two expansions as expansions of one single non singular underlying solution, as underlined in particular by Boulware and Deser soon after Vainshtein work first appeared \cite{Boulware:1973my}. The vDVZ discontinuity and its possible cure {\it \`a la} Vainshtein reappeared later in more complicated (and probably more realistic) models. Those include the Dvali-Gabadadze-Porrati (DGP in the following) model \cite{Dvali:2000hr} and its sequels \cite{CASCA}, which have attracted a lot of attention in particular as far as application to cosmology is concerned \cite{ced,DEBATE1,DEBATE2,DEBATE1bis,CASCACOS};   the ``degravitation'' models \cite{DEGRAV}, the Galileon and k-Mouflage models \cite{GALIK}, 
 the recent models of Refs. \cite{MGED,deRham:2010gu,Chamseddine:2010ub}, as well as possibly the Ho\v{r}ava-Lifshitz theory \cite{Horava:2009uw} (see e.g. \cite{Mukohyama:2009tp}).
 
 The search for an everywhere non singular asymptotically flat solution featuring the Vainshtein recovery has proven to be a difficult task. E.g. Damour {\it et al.} found in Ref. \cite{Damour:2002gp} by numerical integration of the equations of motion, that singularities always appeared in static spherically symmetric solutions of the kind considered by Vainshtein and concluded that the Vainshtein mechanism was not working. This issue was recently reexamined by us, and in Ref. \cite{usprl} we reported briefly (disagreeing with the results of Ref. \cite{Damour:2002gp}) the numerical discovery of a static, spherically symmetric, asymptotically flat solution of massive gravity showing for the first time the Vainshtein recovery of GR. The aim of this work is to provide more details on this solution and on the way it has been obtained.  
 Before doing so, let us however underline that we do not call for a realistic use of the kind of NLPF theory presented here (see however \cite{Damour:2002wu,Chamseddine:2010ub}). Indeed, such theories are believed to suffer from various deadly pathologies, like the the ``Boulware-Deser ghost'' \cite{Boulware:1973my} and the related strong coupling. Rather, we use the NLPF as a toy model to investigate the issue of the success or failure of the Vainshtein mechanism.

 This paper is organized as follows. In the next section \ref{GENSSSS}, mostly introductory, we first introduce the NLPF we will be looking at (subsection \ref{ACE}), the ansatz considered (subsection \ref{ansatz}), the Vainshtein mechanism in brief (subsection \ref{brief}), and then with more details, together with the so-called Decoupling Limit (DL) that played a crucial r\^ole for the numerical integration of the field equations (subsection \ref{subsection results of DL}). Then (section \ref{SectIII}), we give some analytic insight on the putative solution, starting first from standard perturbation theory (subsection \ref{PERTVAC}) yielding in particular a large distance expansion, presenting later expansions valid at intermediate and small distances (subsections \ref{EXPINV} and \ref{INSIDE}). In this section we also discuss with some details the crucial issue of the number of solutions at infinity, arguing that there exists infinitely many such solutions which are not captured by the standard perturbative expansion (subsection \ref{NUMINF}). In the next section \ref{WFA} we introduce a new approximation able to capture both the Vainshtein recovery and the large distance Yukawa decay. Section \ref{section numerics} is devoted to the detailed presentation of our numerical results. We conclude in section \ref{CCL}. Three appendices contain more technicalities.

\section{Generalities about Static Spherically Symmetric Solutions}
\label{GENSSSS}
\subsection{Action and covariant equations}
\label{ACE}
We will consider the theory defined by the action
\begin{equation}
\label{action}
S=\int d^4 x\sqrt{-g}\ \left(\frac{M_P^2}{2} R_g +L_g\right)  + S_{int}[f,g].
\end{equation}
In the above action,  $R_g$ is the Ricci scalar computed with the metric $g_{\mu \nu}$, $f_{\mu\nu}$ is a non dynamical metric, $M_P$ is a mass scale, and $L_g$ denotes a generic matter Lagrangian with a minimal coupling to
the metric $g$ (and not to the metric $f$), and $S_{int}[f,g]$ is an interaction term with non derivative couplings between the two metrics.
Following the notations of Damour {\it et al.} \cite{Damour:2002ws}, these interaction terms are in the form
\be \label{INT}
S_{int}^{(a)} = -\frac{1}{8} m^2 M_{P}^{2} \int d^{4}x {\cal V}^{(a)} (g,f) \equiv  -\frac{1}{8} m^2 M_{P}^{2} \int d^{4}x \sqrt{-g}V^{(a)} ({\bf g^{-1} f}) 
\ee
with ${\cal V}^{(a)}(g,f)\equiv \sqrt{-g}\;V^{(a)}({\bf g^{-1} f})$ a suitable ``potential'' density. There is much freedom in the choice of these potentials. 
In this paper we will mostly concentrate on the following term  (henceforth called AGS), considered in particular
by Arkani-Hamed {\it et al.}  in Ref. \cite{Arkani-Hamed:2002sp},
\be
S_{int}=-\frac{1}{8} m^2 M_{P}^{2}\int d^{4}x \; \sqrt{-g}\;H_{\mu \nu} H_{\sigma \tau}\left(g^{\mu\sigma}g^{\nu\tau}-g^{\mu\nu}g^{\sigma\tau}\right) \label{S3},
\ee
where $g^{\mu \nu}$ denotes the inverse of the metric $g_{\mu \nu}$, and $H_{\mu \nu}$ is defined by
\ba
H_{\mu \nu} = g_{\mu \nu} - f_{\mu \nu}.\nonumber
\ea
We will also comment  on other forms for the potential.
In the equation above, $m$ is the graviton mass and $M_P$ the reduced Planck mass, given by
\be
M_P^{-2} = 8 \pi G_N,\label{Newton}\nonumber
\ee
in term of the Newton constant $G_N$. The theory described by the action (\ref{action})-(\ref{S3}) is invariant under common diffeomorphisms which transforms the metric as
\be
\label{Ginvg}
\begin{aligned}
g_{\mu \nu}(x) &= \partial_\mu x'^{\sigma}(x) \partial_\nu x'^{\tau}(x) g'_{\sigma \tau}\left(x'(x)\right)\;, \nonumber\\
f_{\mu \nu}(x) &= \partial_\mu x'^{\sigma}(x) \partial_\nu x'^{\tau}(x) f'_{\sigma \tau}\left(x'(x)\right)\;.
\end{aligned}
\ee
A crucial property of the potential (\ref{S3}) is that when $g$ is expanded to second order around the canonical Minkowski metric $\eta_{\mu \nu}$
as $g_{\mu \nu} = \eta_{\mu \nu} + h_{\mu \nu}$ and $f$ has the canonical Minkowski form $\eta_{\mu \nu}$, the potential at quadratic order for $h_{\mu \nu}$ takes the Pauli-Fierz form. The equations of motion, derived from action (\ref{action}), read
\be \label{EQMot}
M_P^2 G_{\mu \nu} =\left(T_{\mu \nu}+ T^g_{\mu \nu}\right),
\ee
where $G_{\mu\nu}$ denotes
the Einstein tensor computed with the metric $g$,
$T_{\mu \nu}$ is the energy momentum tensor of matter fields, and
$T^g_{\mu \nu}$ is the effective energy momentum tensor coming from the variation with respect to the metric $g$ of the interaction term $S_{int}$. {\bf $T^g_{\mu \nu}$} depends non derivatively on both metrics $f$ and $g$ and is defined as usual as
\be \label{DEFTMN}
T^{g}_{\mu \nu}(x) = - \frac{2}{\sqrt{-g}} \frac{\delta}{\delta g^{\mu \nu}(x)} S_{int}(f,g).\nonumber
\ee
A simple, but non trivial, consequence of equations (\ref{EQMot}) is obtained by taking a $g$-covariant derivative $\nabla$ of both sides of the equations; one gets, using the Bianchi identities and the conservation of the matter energy momentum tensor, 
\be
\nabla^\mu T_{\mu \nu} =0 \label{CONSCOV},
\ee
the constraint
\be
\nabla^\mu T_{\mu \nu}^g =0 \label{BIAN}
\ee
which the effective energy momentum tensor should obey. This equation will turn to play an important role in the following.

\subsection{Static spherically symmetric ansatz and boundary conditions}
\label{ansatz}
In this paper we study static spherically symmetric solutions of the theory (\ref{action})-(\ref{S3}) and we take the non-dynamical metric $f$ to be flat, i.e. to parameterize (possibly only part of) a Minkowski space-time. Note that it is possible to consider $f$ to be dynamical as well (see e.g. \cite{NONDIAG,Damour:2002wu,Damour:2002ws,Damour:2002gp}), for simplicity we will not consider this possibility here, and consider the flatness of $f$ as a prerequisite of our model. 
Hence, we consider static spherically symmetric configurations using  the following ansatz
\be
\begin{aligned} 
g_{\mu \nu}dx^\mu dx^\nu &= -e^{\nu(R)} dt^2 + e^{\lambda(R)} dR^2 + R^2 d\Omega^2  \; ,\\
f_{\mu \nu}dx^\mu dx^\nu &= -dt^2 + \left(1-\frac{R \mu '(R)}{2}\right)^2 e^{-\mu(R)} dR^2 + e^{-\mu(R)}R^2 d\Omega^2\; ,
\end{aligned}
\label{lammunu}
\ee
where $\nu$, $\lambda$, $\mu$ are unknown functions and $d\Omega^2$ is the metric of a unit 2-sphere 
\be
d\Omega^2 = d \theta^2 + \sin^2\theta d\varphi^2 .\nonumber
\ee 
 This ansatz, where both metrics are diagonal in a common gauge, is not the most general one (see \cite{NONDIAG}) however, it is the one relevant for the study of the Vainshtein mechanism of massive gravity which is the aim of the present work (see e.g. \cite{Damour:2002gp}). 
It is easy to check that for the ansatz (\ref{lammunu}), the tensors $G_{\mu \nu}$, and $T^g_{\mu \nu}$ are both diagonal (the diagonal character of 
$T^g_{\mu \nu}$ follows from the fact its components are made by taking products of the diagonal matrices
${\bf f}$, ${\bf f^{-1}}$, ${\bf g}$ and ${\bf g^{-1}}$). The $tt$ and $RR$ components of (\ref{EQMot}) and the non-trivial part of Bianchi identities (\ref{BIAN}) read
\bse
\label{EOMFULL}
\begin{align}
e^{\nu-\lambda}\left(\frac{\lambda'}{R} + \frac{1}{R^2}(e^\lambda-1)\right) &= 8 \pi G_N\left(T^g_{tt}+\rho e^\nu\right), \label{EOMFULL1}\\ 
\frac{\nu'}{R}+\frac{1}{R^2}\left(1-e^\lambda\right) &= 8 \pi G_N\left(T^g_{RR}+Pe^\lambda\right),\label{EOMFULL2}\\
-\frac{1}{m^{2}M_{P}^{2}}\;\frac{1}{R}\nabla^\mu T_{\mu R}^g &= 0,\label{BIANCHIFULL}
\end{align}
\ese
where the source energy momentum tensor $T_{\mu}^{\;\nu}$ is assumed to have the perfect fluid form 
\be
T_{\mu}^{\;\nu}=\text{diag} (-\rho, P,P,P),\nonumber
\ee 
with total mass 
\be
M\equiv\int_{0}^{R_{\odot}}4\pi R^{2}\;\rho\;dR.\nonumber
\ee
The matter conservation equation reads 
\be\label{MATTCONS}
P'=-\frac{\nu'}{2}(\rho+P).
\ee
In the  integration of the field equations, we will only consider cases where the source has a constant density inside its radius.
For the potential (\ref{S3}), we have
\be
\label{EMT}
T^g_{tt} = m^2 M_P^2\;f_t,\quad 
T^g_{RR} = m^2 M_P^2\;f_R,\quad
\nabla^\mu T_{\mu R}^g = -m^2 M_P^2 f_g,\nonumber
\ee
where we use the following notations \cite{Damour:2002gp}
\bse
\label{f}
\begin{align}
\label{ft}
f_t&= \frac{e^{-\lambda-2 \mu}}{4}\\
	&\quad\times \left[\left(3 e^{\mu+\nu}+e^{\mu}-2 e^{\nu}\right)\left(1-\frac{R \mu'}{2} \right)^2
	+e^{\lambda} \left(2 e^{\mu}-e^{\nu}\right)-3 e^{\lambda+\mu} \left(2 e^{\mu+\nu}+e^{\mu}-2 e^{\nu}\right)
   	\right],\nonumber\\
\label{fR}
f_R&= \frac{e^{-\nu-2 \mu}}{4} \\
	&\times \left[\left(3 e^{\mu+\nu}-e^{\mu}-2 e^{\nu}\right)\left(1-\frac{R \mu'}{2} \right)^2
	+e^{\lambda} \left(2 e^{\mu}+e^{\nu}\right)-3 e^{\lambda+\mu} \left(-2 e^{\mu+\nu}+e^{\mu}+2 e^{\nu}\right)
   	\right],\nonumber\\
\label{fg}
f_g &= -\left(1-\frac{R \mu '}{2}\right)\frac{ e^{-\lambda -2 \mu -\nu }}{8 R}\\
	&\quad\times\left[8 \left(e^{\lambda }-1\right) \left(3 e^{\mu +\nu }-e^{\mu }-e^{\nu }\right)
	+2 R \left(\left(3 e^{\mu +\nu }-2 e^{\nu }\right) \left(\lambda '+4 \mu '-\nu '\right)-e^{\mu} \left(\lambda '+4 \mu '+\nu '\right)\right)\right.\nonumber\\
	&\quad\left.-R^2 \left(\left(3 e^{\mu +\nu }-2 e^{\nu }\right) \left(\lambda' \mu'-2 \mu''-\mu' \nu'+\left(\mu '\right)^2\right)-e^\mu \left(\lambda' \mu'-2 \mu''+\mu' \nu'+\left(\mu'\right)^2\right)-2 e^{\nu } \left(\mu '\right)^2\right)\right].\nonumber
\end{align}
\ese
Besides the components (\ref{EOMFULL1}) and (\ref{EOMFULL2}), the field equations (\ref{EQMot}) have non trivial $\theta \theta$ and $\varphi \varphi$ components. However, it is easy to check that the set of equations (\ref{EOMFULL}), (\ref{MATTCONS}) is equivalent to equations (\ref{EQMot}) and (\ref{CONSCOV}) once the ansatz (\ref{lammunu}) is chosen. 

The system (\ref{EOMFULL}), (\ref{MATTCONS}) is a system of ODEs that requires five boundary conditions (respectively the values of $\nu$, $\lambda$, $\mu$, $\mu'$ as well as $P$ at some (finite) boundary) to be integrated along the radial coordinate $R$. One is also looking for an asymptotically flat solution such that $\nu$ and $\lambda$ must vanish at large $R$, and one should also require that the solution is non singular in $R=0$. To fulfill this condition and avoid 
a conical singularity at the origin, we will require that 
\bea
\lambda(R=0)&=&0 \label{condition on lambda at the origin}.
\eea
The solution found will also be such that 
\be
\mu'(R=0)=0.\nonumber
\ee

In the following subsections we first recall some properties of the so-called Vainshtein mechanism which, if valid, opens a way to recover solutions of General Relativity in massive gravity  for small graviton mass. Then we recall some of our previous results on the so-called decoupling limit of the theory considered, which played a crucial role for the integration of the field equations.

\subsection{Vainshtein mechanism in brief}
\label{brief}
The (quadratic) Pauli-Fierz theory
is known to suffer from the van Dam-Veltman-Zakharov (vDVZ) discontinuity, i.e. the fact that when one lets the mass $m$ of the graviton vanish, one does not recover predictions of General Relativity. E.g., if one adjusts the parameters (namely the Planck scale) such that the Newton constant agrees with the one measured by some type of Cavendish experiment, then the light bending as predicted by Pauli-Fierz theory (and for a vanishingly small graviton mass) will be $3/4$ of the one obtained by linearizing GR \cite{vanDam:1970vg}\footnote{The fact it is smaller is easy to understand: the essential difference between Pauli-Fierz theory and linearized GR comes from an extra propagating scalar mode present in the massive theory. This mode exerts an extra attraction in the massive case compared to the massless case. Hence, if one wants measurements of the force exerted between non relativistic masses to agree, the coupling constant of the massive theory should be smaller than that of the massless theory. But light bending is blind to the scalar sector - because the light energy momentum tensor is traceless. Hence, provided the two theories agree on the force between non relativistic probes, the massive theory would predict a smaller light bending than the massless one.}.  One way to see this is to consider solutions of equations of motion (\ref{EQMot}) which are static and spherically symmetric and which would describe the metric around a spherically symmetric body such as the Sun. To do so, using the ansatz (\ref{lammunu}) is especially convenient because in this form, the $g$ metric can be easily compared with the standard Schwarzschild solution. If one  tries to find a solution expanding in the Newton constant {\bf $G_N$}, as we recall in the next subsection, one finds immediately the vDVZ discontinuity appearing in the form of a different ($m$ independent) absolute value of the coefficients in front of the first non trivial correction to flat space-time in $g_{tt}$ and $g_{RR}$ components  (neglecting the Yukawa decay by assuming the Compton wavelength of the graviton is much larger than other distances of interest). Indeed, those corrections are obtained at first order in $G_N$, i.e. by linearizing the field equations, in which case the NLPF and PF theories are equivalent by definition. However, as first noticed by Vainshtein \cite{Vainshtein:1972sx}, the computation of the next order correction in the NLPF shows that the first order approximation ceases to be valid at distances to the source smaller than a composite scale, the Vainshtein radius defined by 
\ba \label{DEFVAIN}
R_V=\left(m^{-4}R_{S}\right)^{1/5},\nonumber
\ea
where $R_S$ is the Schwarzschild radius of the source. 
This Vainshtein radius obviously diverges when one lets $m$ go to zero. 
It is also much larger than the solar system size for a massive graviton with a Compton wavelength of the order of the Hubble radius and $R_S$ taken to be the Schwarzschild radius of the Sun. Hence, one can not rule out massive gravity based on solar system observations and on the results of the original works on the vDVZ discontinuity \cite{vanDam:1970vg}. Vainshtein also showed that an expansion around the standard Schwarzschild solution (defined as an expansion in the graviton mass $m$) can be obtained (as we recall in the next subsection). This expansion is well behaved when the mass of the graviton is sent to zero, opening the possibility of a recovery of GR solution at small distances $R$ of the source. Indeed, the domain of validity of this second expansion was shown to be $R \ll R_V$. Moreover, the correction found by Vainshtein to the Schwarzschild solution are non analytic in the Newton constant which can explain the failure of the attempt to obtain an everywhere valid solution expanding in the Newton constant.

\subsection{The Decoupling Limit and the Vainshtein mechanism}
\label{subsection results of DL}
The Vainshtein mechanism can be enlightened taking the so-called Decoupling Limit (henceforth DL) of the theory considered.  Indeed this limit gives the scalings of the different metric coefficients expected to hold (for an interesting range of distances to be specified below), should the Vainshtein mechanism be valid. The DL has been investigated by us in a previous work \cite{us}, and those investigations turned also to be crucial to obtain the solutions of the full system presented here and in Ref. \cite{usprl}. Hence, it is the purpose of this subsection to briefly present the salient features of the DL and its application to study of the Vainshtein mechanism. Instead of the way it was originally introduced \cite{Arkani-Hamed:2002sp}, we will define here the DL directly in the equations of motion \cite{us}. Indeed, it can be shown that the full system of equations (\ref{EOMFULL}) reduces to a much simpler one in the DL defined as follows
\be
\label{DEFDEC}
\begin{aligned}
M_{P}&\rightarrow \infty, \\
m & \rightarrow  0,  \\
\Lambda &\sim {\rm constant}, \\
T_{\mu \nu}/M_P & \sim {\rm constant}, \nonumber
\end{aligned}
\ee
where $\Lambda$ is an energy scale defined by 
\ba
\Lambda = \left(m^4 M_P \right)^{1/5}.\nonumber
\ea
This scale is associated with the strongest self interaction of the model (for more details see \cite{Arkani-Hamed:2002sp,us}).
Taking the DL, the equations (\ref{EOMFULL}) become
\bse
\label{EOMDL}
\begin{align}
\frac{\lambda'}{R}+\frac{\lambda}{R^{2}}&=-\frac{m^2}2(3\mu+R\mu') + \frac{\rho}{M_P^2}\label{EOMDL1} \\
\frac{\nu'}{R}-\frac{\lambda}{R^2}&=m^2\mu \label{EOMDL2} \\
\frac{\lambda}{R^2}&=\frac{\nu'}{2R}+Q(\mu) \label{EOMDL3},
\end{align}
\ese
where $Q$ is the only nonlinear part left over in the DL. 
As seen above, the DL amounts to linearizing $G_{tt}^g$ and $G_{RR}^g$ (as well as the sources), keeping the part of $f_t$ and $f_R$ linear in $\mu$, linearizing $f_g$ and keeping also the piece of $f_g$ which is quadratic in $\mu$. As appears above, the linearization of $f_g$ does not contain any pieces depending on $\mu$. $Q$ is the only part of the pieces left over in the DL which depends on the choice of potential $V^{(a)}$ (provided the theory is in the NLPF category). For AGS potential, $Q$ is given by
\be
\label{Q}
Q=-\frac12\left(\frac{\mu'^2}{2}+\mu\mu''+\frac{4\mu\mu'}{R}\right).
\ee
Note that we have neglected the pressure $P$ in the above equations (\ref{EOMDL}), since this is a direct consequence of taking the decoupling limit in the matter conservation equation (\ref{MATTCONS}).
From the three  equations (\ref{EOMDL}) a single equation on $\mu$ can be extracted, 
which after integration and matching of the solution to the source, reads
\bse
\label{mueq}
\begin{align}
2Q(\mu)+\frac{3}{2}\;m^2\mu&=\frac{R_S}{R^{3}}, \qquad \text{(outside source)}, \label{mueq1}\\
2Q(\mu)+\frac{3}{2}\;m^2\mu&=\frac{R_S}{R_\odot^{3}}, \qquad \text{(inside source)} ,\label{mueq2}
\end{align}
\ese
where $R_\odot$ is the radius of the source and where we assumed that the density of the source is constant. Note the r.h.s. of (\ref{mueq2}) is constant,
in contrast to that of Eq.~(\ref{mueq1}). 

The essence of the Vainshtein mechanism, as well as the related scalings of the solutions can be understood from equation (\ref{mueq1}), which allows to obtain $\mu$, as well as from equations (\ref{EOMDL1}) and (\ref{EOMDL2}) which enables to get $\lambda$ and $\nu$ once $\mu$ is known. Indeed, one can guess the existence of  two different regimes for the equation (\ref{mueq1}), 
namely the linear and the nonlinear regimes.
In the linear regime, the nonlinear part $Q$ can first be neglected. This regime is valid for $R\gg R_V$, and the asymptotic solutions, $\bm_{\infty}$ and $\bl_{\infty}$ read
\be
\label{DLsolinf}
\bm_{\infty}\simeq\frac{2}{3}\frac{R_S}{m^2R^{3}}\left[1+{\cal O}\left(\frac{R_V^5}{R^5}\right)\right],\quad
\bl_{\infty}\simeq-\frac{\bn_\infty}2\simeq\frac{2}{3}\frac{R_S}{R}\left[1+{\cal O}\left(\frac{R_V^5}{R^5}\right)\right].
\ee
From (\ref{DLsolinf}), it follows  that $\bn_{\infty}\simeq -2\bl_{\infty}$, which is a manifestation of the vDVZ discontinuity, and results, e.g., in the mentioned difference between light bending in GR and in PF theory. In the nonlinear regime, the linear term in the l.h.s. of (\ref{mueq}) is neglected compared to the nonlinear $Q$-term.
In this regime (valid for $R\ll R_V$) two different scalings can be identified.

The first scaling, henceforth called \emph{Vainshtein scaling}, is found assuming that the l.h.s. cancels the r.h.s. for the leading term in the expansion of $\mu$. I.e., in vacuum, one assumes that the leading term $\bm_{0}$ is such that
\be\label{equation Q Vainshtein regime}
 Q(\bm_{0})\simeq R_S/R^{3}.
 \ee
For $Q$ given by (\ref{Q}), the leading behaviours of the metric function read
\be
\label{DLsolV}
\begin{aligned}
\bm_{V} &\simeq \sqrt{\frac{8R_{S}}{9R}}\left[1+{\cal O}\left(\frac{R}{R_V}\right)^{5/2}\right],\\
\bn_{V} &\simeq -\frac{R_S}{R}\left[1+{\cal O}\left(\frac{R}{R_V}\right)^{5/2}\right],\\
\bl_{V} &\simeq \frac{R_S}{R}\left[1+{\cal O}\left(\frac{R}{R_V}\right)^{5/2}\right].
\end{aligned}
\ee
In this regime, one recovers at dominant order the results of linearized General Relativity, namely $\bn_{V} \sim - \bl_{V}$.
Inside the source, following a similar logic, one can look for a solution to the equation (\ref{mueq2}) $\bm_{\odot,V}$, such that the leading behaviour gives $Q(\bm_{\odot,V})\sim R_S/R_\odot^{3}$.
We find 
\bse
\label{DLsolV in}
\begin{align}
\bm_{\odot,V}& = \sqrt{\frac{R_S}{R_\odot}}\left(\tilde{B}_V-\frac{1}{10\tilde{B}_V}\left(R/R_\odot\right)^2\right)
+\frac{3}{20}\left(mR\right)^2+...\;,\label{DLsolV in1}\\
\bn_{\odot,V}& = \bn_0+\frac{R_SR^2}{2R_\odot^3}+\frac{\tilde{B}_V}4\sqrt{\frac{R_S}{R_\odot}}\left(mR\right)^{2}+...\;,\label{DLsol in2}\\
\bl_{\odot,V}& = \frac{R_SR^2}{R_\odot^3}-\frac{\tilde{B}_V}2\sqrt{\frac{R_S}{R_\odot}}\left(mR\right)^{2}+...\;.\label{DLsol in3}
\end{align}
\ese
where $\tilde{B}_V$ and $\bar{\nu}_0$ are constant of integration, to be fixed by matching the solutions outside the source, Eq.~ (\ref{DLsolV}), and 
inside the source. 

A second scaling is possible, leading to the restoring of General Relativity in the non linear regime \cite{us}.
This is what we called the {\it $Q$-scaling}, whose leading behaviour corresponds to a zero mode of the nonlinear operator $Q$. 
This zero mode is defined such that
$Q(\bm)\sim0$, in which case the r.h.s. term $R_{S}/R^{3}$ compensating the nonlinear piece of (\ref{mueq1}) is subdominant.
The $Q$-scaling solution of (\ref{mueq1}) reads in vacuum,
\bse
\label{DLsolQ}
\begin{align}
\bm_{Q} &= \left(m R_V \right)^{2}\left\{ A_Q \left(\frac{R_V}{R}\right)^2 +
\left[B_Q- \frac{1}{3 A_Q}\ln\left(\frac{R}{R_V}\right)\right]\frac{R}{R_V} \right\}, \label{DLsolQ1}\\
\bn_{Q}&=-\frac{R_{S}}{R}+\frac12 A_Q (R_V m)^4\ln\left(R/R_V\right), \label{DLsolQ2}\\
\bl_Q&=\frac{R_{S}}{R}-\frac{A_Q}2 (R_V m)^4.\label{DLsolQ3}
\end{align}
\ese
From (\ref{DLsolQ}) it is clearly seen that General Relativity is restored for small $R$.
Note that the $Q$-solution (\ref{DLsolQ}) is a richer family compared to the Vainshtein scaling (\ref{DLsolV}),
since it contains two constants, $A_Q$ and $B_Q$. Inside a source, it is not hard to guess that the leading behaviour of the $Q$-scaling is also given by the $Q$-scaling (\ref{DLsolQ}) 
outside the source
\be
\label{DLsolQ in}
\bm_{\odot,Q} = \left(m R_V\right)^2 \tilde{A}_Q \left(\frac{R_V}{R}\right)^2,\qquad 
\bn_{\odot,Q} = -\frac{R_S}{R},\qquad
\bl_{\odot,Q} = \frac{R_S}{R},
\ee
since the leading term of $Q$-scaling does not touch the r.h.s. of Eqs.~(\ref{mueq}),
Note, however, that the subdominant terms are different for the $Q$-scalings inside and outside the source. 
Note also that this scaling leads leads to a curvature singularity at $R=0$ and does not obey the boundary condition (\ref{condition on lambda at the origin}).

In Ref. \cite{us}, we integrated numerically the DL system (\ref{EOMDL}). We found that solutions interpolating between the above described asymptotic regimes indeed exist. 
We were able to identify the following two families of solutions.
For a given choice of parameters $R_S$, $m$ and $R_\odot$, there is one solution $\{\bm_{N,V}, \bn_{N,V}, \bl_{N,V}\}$ which interpolates between the ``Vainshtein'' asymptotic scalings given respectively by (\ref{DLsolinf}), (\ref{DLsolV}) and (\ref{DLsolV in}), as follows,
{\renewcommand{\arraystretch}{2.}\renewcommand{\tabcolsep}{0.4cm}
\be\label{table behavior vainshtein solution}
\begin{array}{c|c|c|c}
\hline
R\quad &\quad R\ll R_\odot \quad & \quad R_\odot\ll R\ll R_V \quad & \quad R\gg R_V\\
\hline
\bm_{N,V} \quad &  \text{const} \quad &   \sqrt{8R_{S}/(9R)} \quad & \quad 2R_S/(3m^2R^{3})\\
\bn_{N,V} \quad & \text{const} \quad &   -R_{S}/R\quad & \quad -4R_S/(3R)\\
\bl_{N,V} \qquad &  \text{const} \times R^2 \quad &  R_{S}/R \quad & \quad 2R_S/(3R)\\
\hline
\end{array}
\ee
Note that the Vainshtein solution is unique for fixed $R_S$, $m$ and $R_\odot$, since the free constant $\tilde{B}_V$ in (\ref{DLsolV in}) is fixed by the requirement of the asymptotic flatness (\ref{DLsolinf})  \cite{us}. Apart from  this Vainshtein solution, we have found a one-parameter family of solutions 
$\{\bm_{N,Q}, \bn_{N,Q}, \bl_{N,Q}\}$ which interpolates between the solutions given by (\ref{DLsolinf}), and the Q-scalings (\ref{DLsolQ}) and (\ref{DLsolQ in}), such that,
\begin{center}
\begin{tabular}{c|c|c|c}
\hline
$R$ & $R\ll R_\odot$ & $R_\odot\ll R\ll R_V$ & $R\gg R_V$\\
\hline
$\bm_{N,Q}$ & $m^2 R_V^4 \tilde{A}_Q\times R^{-2}$ & $m^2 R_V^4 A_Q\times R^{-2}$ & $2R_S/(3m^2R^{3})$\\
$\bn_{N,Q}$ & $-R_{S}/R$ &  $-R_{S}/R$ & $-4R_S/(3R)$\\
$\bl_{N,Q}$ & $R_{S}/R$ &  $R_{S}/R$ & $2R_S/(3R)$\\
\hline
\end{tabular}
\end{center}
In contrast to the Vainshtein solution, this solution is not uniquely fixed by the asymptotic flatness. There is a freedom in the choice of $A_Q$ in (\ref{DLsolQ}), so that the family of solutions corresponds to a curve in the $(A_Q,B_Q)$-plane of solutions (\ref{DLsolQ}). 

Although the solutions we found in  DL  are solutions of a simplified version of the full system, 
they should be valid in some range of distances $R$ for the full system. Indeed, let us estimate the validity of DL solutions (see also \cite{us}). 
To do this we compare the terms left over in the DL system of equations (\ref{EOMDL}) with the terms we dropped from the original system (\ref{EOMFULL}). First of all, keeping only linear terms in $\lambda$ and $\nu$ on the r.h.s. of (\ref{EOMFULL}) requires $R\gg R_S$ outside the source, as in GR.  The second condition comes from comparison of the linear terms containing $\mu$ with those containing $\lambda$ and $\nu$ in the expansion of the r.h.s. of (\ref{EOMFULL1}) and (\ref{EOMFULL2}). In the DL, the linear terms containing $\mu$ are neglected,
and one can see that those terms can only be neglected at distances $R\ll  m^{-1}$ for the full system. 
I.e. those two first conditions state that the DL can at best recover linearized GR and will not capture the expected Yukawa decay of the solution. 
The third condition arises from the neglecting of terms in the nonlinear regime. 
Indeed, the quadratic terms in $\mu$, $\nu$ and $\lambda$ in the r.h.s. of the expansion of (\ref{EOMFULL1}) and (\ref{EOMFULL2}) must be subdominant for DL to be valid for the full system.
Similarly, the cubic terms on r.h.s. of (\ref{BIANCHIFULL}) should be small compared to the quadratic ones, left over in DL. 
This gives nothing new for the Vainshtein scaling, namely, $R\gg R_S$. However, this leads to a more rigid constraint for the $Q$-scaling
\be\label{radius singularity}
R\gg R_V \left(m R_S\right)^{1/5}.
\ee

To summarize, the DL solution, having the Vainshtein asymptotic at small $R$ and the linear asymptotic solution for large $R$, is expected to be a good approximation for the solution of the full system in the range,
\be
R_S\ll R\ll m^{-1},\nonumber
\ee
while the DL solution, approximating the $Q$-scaling at small $R$ and the linear asymptotic solution for large $R$, should be a good approximation for the solution of the full system in the range,
\be
R_V \left(m R_S\right)^{1/5}\ll R\ll m^{-1}.\nonumber
\ee
If a not too compact source of radius $R_{\odot}\gg R_{S}$ is included into the picture, one can check using the dominant behaviour (\ref{DLsolV in}) that the Vainshtein solution even remains a good approximation of a full solution until the origin! This will later be confirmed by our numerical integration. On the other hand, even when a large source is included, the DL $Q$-scaling solution stops being relevant for the full system at the distance $R= R_V \left(m R_S\right)^{1/5}$.

In the following, we will look for a solution of the full system (\ref{EOMFULL}), and we will use our understanding of the DL as a guideline. As we will report, our numerical attempts to find a solution of the full system with the $Q$-scaling asymptotic only lead to singular solutions, and accordingly, we will mostly focus on the extension of the Vainshtein scaling to the full system, rather than on the $Q$-scaling solution.

\section{Integration of the field equations: Analytic insights}\label{SectIII}
Before turning to present the numerical integration of the full system of equations  (\ref{EOMFULL}) and (\ref{MATTCONS}), it is convenient to present below some results one can obtain analytically. Those will mostly be given in the form of series expansions. We have  no proof of the convergence of those series, and we in fact believe (and will give some arguments below) that those series are divergent, being then asymptotic expansions. However, we found those series useful for three main reasons: First they allowed us to set boundary or ``initial'' conditions for the numerical integration with the required precision; second, they were also useful to check that the numerical solutions we found was not singular at the origin of the radial coordinate system; and third, because they help investigating the uniqueness of the solutions at infinity\footnote{Similar series expansions were used in Ref. \cite{Damour:2002gp} to argue that the system we investigate here has a unique asymptotically flat solution. It was then argued on the basis of this, that it was ``a priori'' unlikely that one could match this unique solution to a solution obtained {\it \`a la} Vainshtein, expanding around the Schwarzschild solution, as will be explained in section \ref{EXPINV}.  
As we will see our results do not support this conclusion on the uniqueness of the asymptotically flat solution, hence removing the argument given in Ref. \cite{Damour:2002gp} to explain the singularities found by the numerical integration done in this last reference.}.

Let us first outline the organization of this section as well as its main results, allowing a reader not interested in technical details, to skip all but this introduction. 
The first subsection \ref{PERTVAC} explains with some details the solution of standard perturbation theory. This perturbation theory is defined as an expansion in a small parameter $\epsilon$ where at each order, all the metric function $\lambda, \nu$ and $\mu$ are assumed to be proportional to $\epsilon$ raised to {\it the same} power. The parameter $\epsilon$ is expected to be proportional to the mass of the source (or to its density).  Solutions of the perturbation theory out of the source are obtained at each order in $\epsilon$ in the form of asymptotic series both for small distances and for large distances. Those series are  expressed in terms of the dimensionless quantity $z = R m$. Hence, we obtain a double series expansion (into $\epsilon$ and $z$). The ``small distance'' expansion, which is an expansion valid at small $z$ (at a fixed order $\epsilon$), is similar to the regime (\ref{DLsolinf}) of the DL, and is found (by comparing successive orders in $\epsilon$) to be no longer valid at distances smaller than the Vainshtein radius in agreement with Vainshtein's computation. 
This ``small distance'' series expansion depends on two unknown integration constant which have to be fixed by matching to the source. On the other hand, the ``large distance'' (large z) expansion is uniquely fixed and has no free parameter. However, we argue in subsection \ref{NUMINF} that this is not enough to define a unique solution at infinity, namely because we show that a non perturbative correction can always be added to the unique solution of perturbation theory we identified. This is argued to be related to the fact the series presented are not convergent series, in analogy with what we have shown for the decoupling limit \cite{us}. Then, we present expansions valid inside the Vainshtein radius first outside the source (\ref{EXPINV}) then inside it (\ref{INSIDE}).  The expansion inside the Vainshtein radius (\ref{EXPINV}), but outside of the source, is obtained expanding the functions around GR, in a small parameter $\epsilon'$ which is by definition proportional to $m^2$. At each order in $\epsilon'$, one can then expand the functions in powers of $R/R_S$ ($R_S$ being the Schwarzschild radius of the source). One thus obtain again a double expansion. This expansion,  allows to compute the first non trivial corrections to the GR solution. Those corrections will then later be compared to the ones found by the numerics. 
One key property of the Vainshtein mechanism is related to the existence of those double expansions, to the fact the associated series appear not to be convergent allowing the addition of ``non perturbative corrections'' 
 such as the ones mentioned above, and to the fact, as discovered by Vainshtein, that the solution in the ``below $R_V$'' regime features corrections to GR which are non analytic in the $\epsilon$ parameter of the standard perturbation theory.
Last (\ref{INSIDE}), we obtain a new expansion valid inside the source, that will be later compared to the numerical solution, and also used by us to study the regularity of the solution at the origin.

\subsection{Standard perturbation theory in vacuum}
\label{PERTVAC}
Looking for a solution of the system (\ref{EOMFULL}), (\ref{MATTCONS}) far from a source, a first obvious way is to look for a scheme where the functions  $\lambda$, $\nu$ and $\mu$ can then  be expanded as
\be
\label{dvt l n m chapitre SSS}
\begin{aligned}
\lambda&=\lambda_{0}+\lambda_{1}+... \\
\nu&=\nu_{0}+\nu_{1}+...\\
\mu&=\mu_{0}+\mu_{1}+... 
\end{aligned}
\ee
where $\lambda_{i},\nu_{i},\mu_{i}$ are assumed to be proportional to  $\epsi^{i+1}$, $\epsi$ being a small parameter, and such that $\lambda_{i+1 } \ll \lambda_{i}$, $\nu_{i+1 } \ll \nu_{i}$ and $\mu_{i+1 } \ll \mu_{i}$, at least for some range of distances. This is what we will call here standard perturbation theory. Before  going further, let us first introduce a new dimensionless  radial coordinate $z$ defined by $z=Rm$. In this unit, $R=m^{-1}$ corresponds to $z=1$, and $R=R_{V}$ translates to $z=a$, where $a$ is defined as 
\be\label{def a}
a \equiv R_{V}m.
\ee
Using this variable, the lowest order $\lambda_{0},\nu_{0}$ and $\mu_{0}$ are obtained solving the system (\ref{EOMFULL1}-\ref{BIANCHIFULL}) linearized which reads 
\be
\label{linsystem}
\begin{aligned}
\frac{\lambda'}{z}+\frac{\lambda}{z^2}+\frac12\left(\lambda+3\mu+z \mu'\right)&=0,\\
\frac{\nu'}{z}-\frac{\lambda}{z^2}-\frac12\left(\nu+2 \mu\right)&=0,\\
\nu'-\frac{2\lambda}{z}&=0.\nonumber
\end{aligned}
\ee
The asymptotically flat solution of the above linear system can be found analytically, it is given by \cite{Aragone:1972fn,Damour:2002gp} 
\bse
\label{linsolution}
\begin{align}
\nu_0&= -\mathcal{C}\;\frac{4b}{3z}e^{-z},\\
\lambda_0&= \mathcal{C}\;\frac{2b}{3}\left(1+\frac1z\right)e^{-z},\\
\mu_0&= \mathcal{C}\;\frac{2b}{3z}\left(1+\frac1z+\frac{1}{z^2}\right)e^{-z},
\end{align}
\ese
where 
\ba\label{def b}
b \equiv  a^5 = (R_{V}  m)^{5} = R_{S} m,\nonumber
\ea
and $\mathcal{C}$ is a dimensionless integration constant.  The value of the constant of integration $\mathcal{C}$ can in principle be determined by integrating the system inwards and matching to the source, as can be done in a similar case in GR. There, indeed, an integration constant with the same status arises, when one integrate the linearized Einstein equations in the vacuum around a static spherically symmetric source. This constant of integration is then seen to give the Schwarzschild mass, once matched to the source. Here, an alternative way to proceed, is to use our knowledge of the Decoupling Limit. Indeed, requiring 
 the solution to be very close to the Decoupling Limit solution  (\ref{DLsolinf}) in the range $R_{V}\ll R\ll m^{-1}$, \emph{i.e.} $a\ll z\ll 1$, fixes  the constant to be 
\be\label{def C}
\mathcal{C}=1.
\ee
In Section \ref{subsection numerical results}, we will show that this choice of constant of integration is also in agreement with the numerical integration. Accordingly,  from here-on and until the end of this article, we will assume (except at some scattered places)} $\mathcal{C}=1$. Note however, that when dealing with the equations in vacuum, the value of this constant is irrelevant.

Having fixed ${\cal C}$ one notices that the solution (\ref{linsolution}) features the expected behaviour for the gravitational potential of a massive graviton: it has a Yukawa decay and the ``Newtonian'' part $\nu_0$  is $4/3$ larger than the GR result (in the range of distances $R \ll m^{-1}$) while the radial part $\lambda_{0}$ is $2/3$ smaller, a structure related to the vDVZ discontinuity.  Our expansion parameter $\epsi$, can then be defined as $\epsi \equiv {\cal C} \times b$, note that it is indeed much smaller than one for usual astrophysical sources and a graviton of cosmological size Compton length.

To go beyond the lowest order (\ref{linsolution}), the first step 
consist in re-writing the system of equations (\ref{EOMFULL}), separating explicitly the linear terms from the nonlinear ones: 
\be
\label{expansion separe lin quad}
\begin{aligned}
\frac{\lambda'}{z}+\frac{\lambda}{z^{2}} + \frac{1}{2}(\lambda+3\mu+ z \mu') &=f_{t,\geqslant2}-G_{tt,\geqslant2}\;,\\
\frac{\nu'}{z}-\frac{\lambda}{z^{2}}-\frac{1}{2}(\nu+2\mu) &= f_{R,\geqslant2}-G_{RR,\geqslant2}\;,\\
\frac{\lambda}{z^{2}}-\frac{\nu'}{2z} &= \frac{1}{z}f_{g,\geqslant2}.
\end{aligned}
\ee
In the above system, we have used the expression (\ref{f}) for the functions $f_{i}$ 
and the $f_{i,\geqslant2}$, and $G_{ii,\geqslant2}$ stand for the nonlinear part of the functions $f_{i}$ and of the components of the Einstein's tensor in spherical coordinates. At each order $n$, the functions $\lambda_{n},\nu_{n}$ and $\mu_{n}$ (with $n \geq 1$) can be obtained by solving the above system where the left hand side consists in a linear operator acting on the unknown $\lambda_{n},\nu_{n}$ and $\mu_{n}$, while the right hand side is obtained keeping the order $\epsi^{n+1}$ and consists in non linear expressions depending on the $\lambda_{i},\nu_{i}$ and $\mu_{i}$ (with $i < n$). As will be shown in detail in the appendix \ref{APPA}, the functions $\lambda_{n},\nu_{n}$ and $\mu_{n}$ (with $n \geq 1$) can be expressed as asymptotic (and formal) expansions in large $z$ (large distance) or small $z$ (small distance) regimes. We now discuss the main properties of those expansions, starting, for reasons we will explain, by the small distance expansion.

\subsubsection{Small distance expansion}
\label{SDExp}
An expansion, asymptotic at small $z$, can be obtained in the form 
\be 
\label{EXPANSN}
\begin{aligned}
\mu_n &= z^2 \left(\epsi \frac{e^{-z}}{z^5}\right)^{n+1}  \sum_{i=0}^{\infty}\sum_{k=0}^{n} \mu_{n,i,k}\; z^{i}\left(z^5\log z\right)^{n-k}, \\
\lambda_n &= z^4 \left(\epsi \frac{e^{-z}}{z^5}\right)^{n+1}  \sum_{i=0}^{\infty}\sum_{k=0}^{n} \lambda_{n,i,k}\; z^{i}\left(z^5\log z\right)^{n-k}, \\
\nu_n &=  z^4 \left(\epsi \frac{e^{-z}}{z^5}\right)^{n+1}  \sum_{i=0}^{\infty}\sum_{k=0}^{n} \nu_{n,i,k}\; z^{i}\left(z^5\log z\right)^{n-k}, 
\end{aligned}
\ee
Note that the form of this expansion was sketched in Ref. \cite{Damour:2002gp}, with however some important differences: the corresponding expansion given there is stopping at a finite value of $i$ at each order $n$ and did not contain any logarithms. As shown in the appendix \ref{APPA}, it is however not possible to obtain such a simplified expansion. As detailed in the same appendix, two important properties of  the expansion (\ref{EXPANSN}) are as follows. First, when solving for the order $n$, the coefficients $\mu_{n,i,k}, \lambda_{n,i,k}$ and $\nu_{n,i,k}$ are determined only after choosing two arbitrary constants. This is because the linear operator appearing on the left hand side of equation (\ref{expansion separe lin quad}) has two independent zero modes, (among which the zero order solution (\ref{linsolution})) which can be expanded formally in the same way as above\footnote{The presence of logarithms in this expansion is also related to those zero modes.}. One way to fix those constants would be to enforce at small z the hierarchy 
$\lambda_{i+1 } \ll \lambda_{i}$, $\nu_{i+1 } \ll \nu_{i}$ and $\mu_{i+1 } \ll \mu_{i}$. We have in fact checked that it is possible to do so at order $n=1$. However, it is hard to go further because it is far from obvious that the expansions considered are convergent. The question of convergence is actually quite hard to address, since it requires to know the behaviour of the coefficients $\lambda_{n,i,k}$, $\nu_{n,i,k}$ and $\mu_{n,i,k}$. This can be achieved quite easily for a subset of those coefficients.
Indeed, let us, at each order $n$ retain the dominant term in the expansion (\ref{EXPANSN}) at small $z$ (i.e. for $R\ll m^{-1}$). We obtain, for $z\ll 1$ the following expansion for $\mu$ 
\be
\mu \sim \sum_{n=0}^{\infty} \epsi ^{n+1}\frac{\mu_{n,0,n}}{z^{3+5n} }\;.\label{solution LD serie chapitre SSS}
\ee
Equivalently, this is what is obtained in the limit $m\rightarrow 0$. In fact, and this is the second important property we would like to mention here, as shown in appendix \ref{APPA}, the expansion (\ref{solution LD serie chapitre SSS}) is exactly the one obtained in the DL (\emph{cf.} Ref. \cite{us}, Eq. (4.34)). The coefficient $\mu_{n,0,n}$ can be easily computed in the DL, and it has been shown in \cite{us} that this series diverges for any $z$, providing only an asymptotic expansion of the solution. One can even go further: using the exact values of the coefficients $\mu_{n,0,n}$ obtained in the DL, one can show that the subseries of (\ref{EXPANSN}) given by 
\be
\sum_{n=0}^{\infty} \epsi^{n+1} e^{-(n+1)z} \frac{\mu_{n,0,n}}{z^{3+5n} }, \label{solution LD + exponential serie chapitre SSS}
\ee
diverges for any $z$, despite the presence of the exponential damping $e^{-(n+1)z}$. Of course, the fact that this sub-series of (\ref{EXPANSN}) diverges does not lead to any conclusion by itself besides the fact that the expansion (\ref{EXPANSN}) cannot be absolutely convergent (if it were the case, any subseries would converge and one could rearrange at will the order of summation). In our case, there is no clear prescription for the order of summation, and the divergence of the sub-series (\ref{solution LD + exponential serie chapitre SSS}) does not mean that the full series diverges. Still, we see this divergence as a clue that the convergence of the full series is far from guaranteed, and that any conclusion based on these expansions should be taken with great care. For instance, one may question the approach followed in \cite{Damour:2002gp}, Sec. IV.B, which consists in building \emph{the} solution of massive gravity far from the source and arguing on its uniqueness, starting from an expansion of the form (\ref{EXPANSN}), which is implicitly assumed to be convergent. Moreover, as also happens in the  DL, even if the series expansion (\ref{EXPANSN}) turns out to be uniquely defined (once the constants we mentioned have been fixed e.g. by the procedure outlined above), this does not mean that it defines a unique vacuum solution 
 because it could be, if the series is not convergent, that a family of solutions all have the same asymptotic series expansion. We will in fact come back later to this important issue. 

Though, let us state that from a numerical point of view expansions  (\ref{EXPANSN}) and  (\ref{solution LD serie chapitre SSS}) are very useful. Indeed, if one keeps only the relevant first orders, these expansions appear to be excellent approximations of the solutions and can be used to set properly boundary or initial conditions in the numerical integration.

Moreover, by comparing two successive order in $\epsilon$, it is easily seen that for $R \ll m^{-1}$, the expansion (\ref{EXPANSN}) ceases to be valid when $R$ becomes smaller than the Vainshtein radius (i.e. whenever $\epsilon z^{-5}$ becomes of order one).

\subsubsection{Large distance expansion}
In a similar way as above, an asymptotic expansion at large $z$ can be found at each order $n$ in the $\epsi$ expansion. It is derived in appendix \ref{APPA}, and reads 
\be
\label{LARGEZ}
\begin{aligned}
\mu_n &= \epsi^{n+1} e^{-(n+1)z} \sum_{i=-\infty}^{i=0} \mu_{n,i} z^i,  \\
\lambda_n &= \epsi^{n+1} e^{-(n+1)z} \sum_{i=-\infty}^{i=0} \lambda_{n,i} z^i, \\
\nu_n &= \epsi^{n+1} e^{-(n+1)z} \sum_{i=-\infty}^{i=0} \nu_{n,i} z^i.
\end{aligned}
\ee
One can verify easily that the zeroth order exact solutions (\ref{linsolution}) can be expressed as above.
In contrast to the expansion (\ref{EXPANSN}), there is here no need for logarithms nor to fix ``constants of integration''. As such, this could indicate in a clearer way than in the previous case that, as stated in Ref. \cite{Damour:2002gp}, perturbation theory defines a unique solution. This is however again not obvious because this relies crucially on the fact the above series converges. In fact, we have computed numerically the first coefficients of the first non trivial order $n=1$ and check that the corresponding series seems to have a vanishing radius of convergence\footnote{One finds numerically that $\left|\mu_{1,i}/\mu_{1,i+1}\right| \propto|i|$ indicating a vanishing radius of absolute convergence}. In the next section we try to address the issue of the uniqueness of the solution at infinity in a different way. This analysis again indicates, in agreement with our numerical investigations, that the solution at infinity is not unique.

\subsection{Investigations of the number of solutions at infinity}
\label{NUMINF}
As underlined in the previous subsection, it is hard to conclude anything on the uniqueness of the solution at infinity from the existence of series expansions, which can be divergent. Here we would like to provide arguments showing that there are in fact infinitely many asymptotically flat solutions at infinity which have the same dominant behaviour at large distance once a constant, corresponding to the ``mass'' of the source in the dominant terms has been chosen. This non-uniqueness appears in the form of a non perturbative correction that can be added at will to the unique solution of the perturbation 
theory.\footnote{Note that we found similar result in the DL case \cite{us}, as recalled in Sec. \ref{subsection results of DL}. Indeed, in the DL case, these infinitely many solutions allowed to accommodate several types of solutions at small distances: the Vainshtein solution (\ref{DLsolV}) and the Q-scaling solutions (\ref{DLsolQ}) which depend on a free constant $A_{Q}$. These solutions also appear as ``non perturbative''. It is interesting to note that we can arrive to the same conclusion (that there is an infinite number of solutions at the infinity with the same asymptotic behaviour) 
 if we consider the weak-field equation approximation (see below, Sec. \ref{WFA}) as a starting point for this proof. 
The general structure of the additional solutions remains the same as for NLPF (see below in (\ref{finalsolz})), with, however,
different exponent and pre-exponent factors. 
This difference is due to the fact that when getting an analogue of (\ref{lindelta1}) from the weak-field equations 
(\ref{master}), some terms will be missing compared to (\ref{lindelta1}).}
This is in striking contrast with the General Relativity case. Indeed, in GR, once the asymptotic behaviour has been chosen through\footnote{We recall that in GR, the vacuum solution depends on one integration constant, which can be identified with the Schwarzschild radius $R_{S}$.} $\nu=-R_{S}/R$ and  $\lambda=R_{S}/R$, there is a unique solution which follows this asymptotic behaviour at infinity: the  Schwarzschild solution which reads $\nu = \nu_{GR} \equiv \ln(1-R_S/R)$ and $\lambda = \lambda_{GR} \equiv - \nu_{GR}$.
In NLPF Gravity $-$ at least for potentials of the type (\ref{S3}) $-$ we shall see that the asymptotic behaviour is not enough to determine uniquely the solution.

To prove the existence of these infinitely many solutions,  let's first assume that there exists a solution of the full nonlinear system of equations (\ref{EOMFULL}), 
$\bar\sigma\equiv\{\bar\lambda,\bar\nu,\bar\mu\}$, such that 
\be
\bar{\sigma}\to\sigma_0  \quad \text{for} \quad z\to\infty.\nonumber
\label{asympt}
\ee
where $\sigma_0\equiv\{\lambda_0,\nu_0,\mu_0\}$ as defined in Eq.~(\ref{linsolution}), and we imply we have fixed all the necessary integration constants such that $\bar{\sigma}$ and $\sigma_0$ are fully specified.
We then want to show there is an infinite number of solutions  in vacuum having the same asymptotic at the infinity.
We will look for these solutions $\sigma$ as small perturbations of the known solution  $\bar{\sigma}$:
\be
\sigma=\bar\sigma+\delta\sigma,\label{delta}\nonumber
\ee
where $\delta\sigma\equiv\{\delta\lambda,\delta\nu,\delta\mu\}$
and we will further assume that
\be
|\delta\sigma|\ll |\bar\sigma|, \quad |\delta\sigma'|\ll |\bar\sigma'|,\quad |\delta\sigma''|\ll |\bar\sigma''|. \label{assumedelta}
\ee
Taking into account (\ref{assumedelta}), we linearize the system of equations (\ref{EOMFULL}),  for $z\to \infty$,
around the solution $\bar\sigma$ to obtain the following system of equations on $\delta\lambda$, $\delta\nu$ and $\delta\mu$:
\bse
\label{lindelta1}
\begin{align}
\frac{\delta\lambda'}{z}+\frac{\delta\lambda}{z^2}+\frac12\left(\delta\lambda+3\delta\mu+z \delta\mu'\right)&=
	\frac{be^{-z}}{3z}\delta\nu,\\
\frac{\delta\nu'}{z}-\frac{\delta\lambda}{z^2}-\frac12\left(\delta\nu+2 \delta\mu\right)&=0,\\
\frac{\delta\nu'}{2z}-\frac{\delta\lambda}{z^{2}}&=
	\frac{be^{-z}}{3z^{2}}\left(\delta\mu''-\delta\mu'-2\delta\mu+\frac{\delta\lambda'}{z}+2\delta\nu\right).
\end{align}
\ese
Note that we substituted $\sigma_0$ instead of $\bar\sigma$ in (\ref{lindelta1}), since $\bar\sigma\to\sigma_0$ at $z\to\infty$ and 
we are interested in solutions at infinity.

Until now, we have not made any assumptions about the form of $\delta\sigma$, except (\ref{assumedelta}).
If in addition one assumes that the r.h.s. of (\ref{lindelta1}) can be neglected, then the solution for $\delta\sigma$
will be given by (\ref{linsolution}) provided that the integration constant $b$ is replaced by $\delta b$, such that 
$|\delta b|\ll |b|$. 
However in this case one should impose $\delta b=0$ since we require $\sigma\to \sigma_0$ at $z\to\infty$,
thus we do not obtain any new solution (this is, {\it mutatis mutandis}, what happens in GR).

Instead, let us try to find another solution such that (some) terms containing $e^{-z}$ are not negligible in (\ref{lindelta1}). In particular, we would like to keep the term containing $\delta\mu''$.
Introducing  a new variable $\zeta$ as follows,
\be
\zeta\equiv e^z, \label{zeta}\nonumber
\ee
the system of equations (\ref{lindelta1}) can be rewritten as,
\bse
\label{lindelta2}
\begin{align}
&\frac{\zeta\dot{\delta\lambda}}{\log\zeta}+\frac{\delta\lambda}{(\log\zeta)^2}
	+\frac12\left(\delta\lambda+3\delta\mu+\zeta\log\zeta \,\dot{\delta\mu}\right)=
	\frac{b\,\delta\nu}{3\zeta\log\zeta},\\
&\frac{\zeta\dot{\delta\nu}}{\log\zeta}-\frac{\delta\lambda}{(\log\zeta)^2}-\frac12\left(\delta\nu+2 \delta\mu\right)=0,\\
&\frac{\zeta\dot{\delta\nu}}{2\log\zeta}-\frac{\delta\lambda}{\log^{2}\zeta}=
	\frac{b}{3\zeta\log^{2}\zeta}\left(\zeta^2\ddot{\delta\mu}-2\delta\mu+\frac{\zeta\dot{\delta\lambda}}{\log\zeta}+2\delta\nu\right),
\end{align}
\ese
where dot denotes the derivative with respect to $\zeta$. 
In order to simplify the system of equations (\ref{lindelta2}),  
let us make further assumptions about the functions $\delta\lambda$, $\delta\nu$ and $\delta\mu$ at $\zeta\to\infty$:
\be
|\delta\nu|\ll |\delta\mu| \ll \zeta|\dot{\delta\mu}|\ll \zeta^2|\ddot{\delta\mu}| \ll \zeta^3|\dddot{\delta\mu}|,
\quad 
\log\zeta|\delta\lambda| \ll \zeta|\dot{\delta\lambda}|\ll \zeta|\delta\lambda|.
 \label{assumedelta1}
\ee
The equations (\ref{lindelta2}) then take the simpler form
\bse
\label{lindelta3}
\begin{align}
 \dot{\delta\lambda}+\frac12(\log\zeta)^2\dot{\delta\mu}=0,\\
 \dot{\delta\nu}-\frac{\delta\lambda}{\zeta\log\zeta}-\frac{\log\zeta}{\zeta}\delta\mu=0,\\
 \ddot{\delta\mu}+\frac{3}{2b\zeta}\left(2\delta\lambda-\zeta\log\zeta\,\dot{\delta\nu}\right)=0,
 \end{align}
\ese
These equations can be combined to give one differential equation on $\mu$:
\be
\zeta\dddot{\delta\mu}+\ddot{\delta\mu}-\frac{9}{4b}(\log\zeta)^2\dot{\delta\mu} - \frac{3\log\zeta}{b\zeta}\delta\mu=0,\nonumber
\ee
which --- applying the assumptions (\ref{assumedelta1}) --- can be rewritten as
\be
\dddot{\delta\mu}-\frac{9}{4b\zeta}(\log\zeta)^2\dot{\delta\mu}=0.\label{finaleq}
\ee
The asymptotic form of the decaying solution of Eq.~(\ref{finaleq}) can be easily guessed:
\be
\delta\mu=F_\infty(\zeta)\exp\left(-3\sqrt{\frac{\zeta}{b}}\log\zeta\right),\quad \text{for}\quad \zeta\to \infty,\label{finalsol}
\ee
where $F_\infty(\zeta)$ is a  slowly varying function compared to $\exp\left(-\frac{3}{\sqrt{b}}\, z\, e^{z/2}\right)$ that cannot be fixed through this leading behaviour analysis (\emph{cf.} Appendix \ref{Infinitely many solutions at infinity: a series expansion approach} for a more detailed  approach based on series expansions).
Note that a companion growing mode can also be considered
\be
\delta\mu=F^{(grow)}_\infty(\zeta)\exp\left(3\sqrt{\frac{\zeta}{b}}\log\zeta\right),\quad \text{for}\quad \zeta\to \infty,\label{finalsol growing}
\ee
which is fixed to zero by the  conditions (\ref{assumedelta});  this growing mode can play an important r\^ole while integrating the equations of motion {\it\`a la} Runge-Kutta, since it can blow up very rapidly (as exponent of exponent) if sourced by numerical errors. This explains the difficulty one encounters  while trying to integrate the equations of motion for distances beyond $R_{V}$ for the potential (\ref{S3}). In the following, we will discard such a term since it contradicts the boundary conditions at infinity, but one should keep in mind its possible existence when solving the equations numerically.

From (\ref{lindelta3}) one can find the other functions at $\zeta\to\infty$,
\bse
\label{finalsol2}
\begin{align}
\delta\lambda&=-F_\infty(\zeta)\frac{(\log\zeta)^2}{2} \exp\left(-3\sqrt{\frac{\zeta}{b}}\log\zeta\right),\\
\delta\nu&=-F_\infty(\zeta)\frac{\sqrt{b}}{3}\frac{1}{\sqrt{\zeta}} \exp\left(-3\sqrt{\frac{\zeta}{b}}\log\zeta\right).
\end{align}
\ese
Finally, one can check that the assumptions (\ref{assumedelta}) and (\ref{assumedelta1}) are satisfied for the 
solutions (\ref{finalsol}), (\ref{finalsol2}) as $\zeta\to\infty$ ($z\to\infty$). 
In terms of the variable $z$, equations (\ref{finalsol}) and (\ref{finalsol2}) read
\bse
\label{finalsolz}
\begin{align}
\delta\mu&=F_\infty(z)\exp\left(-\frac{3}{\sqrt{b}}\, z\, e^{z/2}\right),\\
\delta\lambda&=-F_\infty(z)\frac{z^2}{2} \exp\left(-\frac{3}{\sqrt{b}}\, z\, e^{z/2}\right),\\
\delta\nu&=-F_\infty(z)\frac{\sqrt{b}}{3} \exp\left(-\frac{z}{2}-\frac{3}{\sqrt{b}}\, z\, e^{z/2}\right).
\end{align}
\ese
Note that it is possible to complete this leading behaviour analysis through a series expansion of the solution;  we refer the reader to Appendix \ref{Infinitely many solutions at infinity: a series expansion approach} for more details.

Thus we have shown that apart from the solution $\bar\lambda$, $\bar\nu$ and $\bar\mu$, there is at least
one family of asymptotic solutions at $z\to\infty$, parametrized by the function $F_{\infty}$, and given by 
\be
\begin{aligned}
\lambda&=\bar\lambda+\delta\lambda,\nonumber\\
\nu&=\bar\nu+\delta\nu,\nonumber\\
\nu&=\bar\nu+\delta\nu.
\end{aligned}
\ee
Note, that the difference between solutions decays extremely fast --- as a double exponent --- as $z\to\infty$.

\subsection{Expansion inside the Vainshtein radius and outside the source}\label{subsection Expansion inside the Vainshtein radius} \label{EXPINV}
Let us now discuss what happens inside the Vainshtein radius, but outside the star.
This will prove useful for different reasons. We will indeed recall how one can obtain an expansion around the GR solution, and hence compute the corrections to GR first found by Vainshtein. This will later be compared in section \ref{subsection numerical results} to the numerical solution. We will also comment on the r\^ole of the integration constants appearing in the expansion procedure. 

Let us first rely on our investigation of the DL, which provides interesting insights into the different types of solutions in the region $R_{\odot}\ll R\ll R_{V}$.
It has been shown in \cite{us} (as also recalled in Sec. \ref{subsection results of DL}) that, in the DL, two types of solutions exist in the region $R_{\odot}\ll R\ll R_{V}$: the so-called Vainshtein solution of Eq. (\ref{DLsolV}) and the Q-scaling solutions (\ref{DLsolQ}). We are interested in obtaining, in the full theory, a solution which will be close to one of these DL solutions, for a range of distances as broad as possible. It has been recalled in Sec. \ref{subsection results of DL} that the $Q$-scaling solutions stops being relevant for the full nonlinear theory at the radius $R_{*}= R_V \left(m R_S\right)^{1/5}$. Below this distance, the DL is no longer useful for understanding the theory: new nonlinear terms enter into the game and one has to rely on numerics to investigate the existence of solutions. We will come back to numerics in Sec. \ref{section numerics}, but we can already announce that our investigation has been negative in this direction: it has not been possible to us to find a regular solution of the full system which would exhibit a $Q$-scaling behaviour in the range $R_{*}=R_V \left(m R_S\right)^{1/5}\ll R\ll R_{V}$: all our numerical $Q$-scalings solutions were developing  strong instabilities around the problematic radius $R_{*}$, and integration was not possible for smaller distances. For this reason, we will mostly focus in the rest of this article on the Vainshtein solution, and try to generalize the DL version of this solution to the full system, under the form of series expansions.
A first resolution method consists in looking for an expansion whose zero-th order is just the DL. A convenient parameter for such an expansion is the parameter $a=R_{V}m$ introduced in Eq. (\ref{def a}). Indeed, one can show  \cite{us} that after a rescaling of the functions and equations, the limit $a\rightarrow 0$ in the system (\ref{EOMFULL}) leads to the DL equations (\ref{EOMDL}). The full solution expansion  reads
\be\label{expansion in a}
\mu=a^{2}\sum_{n=0}^{\infty}\mu^{(a)}_{n}(R/R_{V})\; a^{2n},\quad \lambda=a^{4}\sum_{n=0}^{\infty}\lambda^{(a)}_{n}(R/R_{V})\; a^{2n},\quad \nu=a^{4}\sum_{n=0}^{\infty}\nu^{(a)}_{n}(R/R_{V})\; a^{2n},
\ee
where the zero-th order functions $a^{2}\mu^{(a)}_{0}(R/R_{V}),a^{2}\lambda^{(a)}_{0}(R/R_{V}),a^{2}\nu^{(a)}_{0}(R/R_{V}) $ are the solutions of the DL system, whose expansions are given in Eq.  (\ref{DLsolV}). Even if the expansion (\ref{expansion in a}) has the advantage to link explicitly the solution of the full system to the DL, it does not allow for an easy comparison with the GR solution. For this purpose, another expansion is possible as was originally proposed by Vainshtein \cite{Vainshtein:1972sx}.

Indeed, it is possible to look for a solution as an expansion in powers of the (square of the) graviton mass $m^2$ dealing with the mass term as a perturbation around massless GR.  Let us define this expansion as
\be\label{expansion in m}
\mu=\sum_{k=0}^{\infty}\mu^{(m)}_{k}(R/R_{S})\; m^{2k},\quad \lambda=\sum_{k=0}^{\infty}\lambda^{(m)}_{k}(R/R_{S})\; m^{2k},\quad \nu=\sum_{k=0}^{\infty}\nu^{(m)}_{k}(R/R_{S})\; m^{2k}.
\ee
By definition, it is such that the lowest order   is equal to the Schwarzschild solution
\ba \label{SCHWA}
\lambda^{(m)}_{0}=-\nu^{(m)}_{0}=-\ln \left(1-\frac{R_{S}}{R}\right)=\frac{R_{S}}{R}+\frac{1}{2}\left(\frac{R_{S}}{R}\right)^{2}+...
\ea
Following Vainshtein, let us now briefly sketch how to obtain the first other non trivial terms of the expansion. The function  $\mu^{(m)}_{0}$ can be  found using the Bianchi identity (\ref{BIANCHIFULL}), where $\lambda$ and $\nu$ have been replaced by their GR expression $\lambda^{(m)}_{0},\nu^{(m)}_{0}$:
\be
-\frac{1}{R}f_{g}\left[\lambda_{0}^{(m)},{\lambda_{0}^{(m)}}',\nu_{0}^{(m)},{\nu^{(m)}_{0}}',\mu_{0}^{(m)},{\mu_{0}}^{(m)},{\mu_{0}^{(m)}}''\right]=0
\label{Bianchi ordre 0}.
\ee
One can look for a solution for $\mu_{0}^{(m)}$ as an expansion in powers of $\left(R_{S}/R\right)$, valid far from the source, where $R\gg R_{S}$. In this regime, the functions $\lambda^{(m)}_{0},\nu^{(m)}_{0}$ go to zero, meaning that  it makes sense to linearize the Bianchi equation in $\lambda^{(m)}_{0}$ and $\nu^{(m)}_{0}$ in order to find the dominant behaviour of $\mu_{0}^{(m)}$. In addition, if $\mu_{0}^{(m)}$ is assumed to also tend towards zero, it is consistent to only keep  the terms of lowest order in this function. Since there is no linear term in $\mu_{0}^{(m)}$ in the Bianchi equation, this lowest order is nothing else than the quadratic term $Q$ defined in Eq. (\ref{Q}). The Bianchi equation  (\ref{Bianchi ordre 0})  then reads, at dominant order
\be
\frac{\lambda_{0}}{R^{2}}=\frac{\nu_{0}'}{2R}+Q(\mu_{0}).\label{equation bianchi ordre 0 petites distances}\nonumber
\ee
Using the asymptotic behaviour far from the source of the Schwarzschild solution (\ref{SCHWA}), \emph{i.e.} the Newtonian regime, one gets the equation
\be \label{EQQSIMP}
Q(\mu_0) = \frac{R_S}{ 2 R^3},\nonumber
\ee
which is identical to the DL equation (\ref{equation Q Vainshtein regime}). If one assumes that the function $\mu_{0}^{(m)}$ can be expanded as a power series in $R$, one finds
\ba 
\mu_{0}^{(m)}&\sim& \sqrt{\frac{8R_{S}}{9R}}.\label{SCAbis} 
\ea
The higher powers in $R$ can then be found order by order in $R$  \cite{Damour:2002gp}. Let us now compute the first order terms. The functions $\lambda^{(m)}_{1}$ and $\nu^{(m)}_{1}$ are computed through the Einstein's equations (\ref{EOMFULL1}) and (\ref{EOMFULL2}). In the limit $R_S  \ll R \ll m^{-1}$, these equations read
\ba
\frac{{\lambda_{1}^{(m)}}'}{R}+\frac{\lambda_{1}^{(m)}}{R^{2}}&=&-\frac{m^{2}}{2}(3\mu_{0}^{(m)}+R{\mu_{0}^{(m)}}') \label{lin1bis}\nonumber\\
\frac{{\nu_{1}^{(m)}}'}{R}-\frac{\lambda_{1}^{(m)}}{R^{2}}&=&{m^{2}}\mu_{0}^{(m)}\label{lin2bis}.\nonumber
\ea
Using  the behaviour (\ref{SCAbis}), the leading behaviour of the first order functions can be found to be
\be
\label{Chapitre SSS lSR}
\begin{aligned}
\nu_{1}^{(m)}&=  \left(m R \right)^{2} \left[\frac{2\sqrt{2}}{9}\sqrt{\frac{R_{S}}{R}}+\mathcal{O}\left(\frac{R_{S}}{R}\right)\right]\;,\\
\lambda_{1}^{(m)}&= \left(m R \right)^{2} \left[-\frac{\sqrt{2}}{3}\sqrt{\frac{R_{S}}{R}}+\mathcal{O}\left(\frac{R_{S}}{R}\right)\right]\;, 
\end{aligned}
\ee
while the function $\mu_{1}^{(m)}$ is obtained using the Bianchi identity and  reads
\bea
\mu_{1}^{(m)}&=& \left(m R \right)^{2}\left[\frac{6}{31}+\mathcal{O}\left(\sqrt{\frac{R_{S}}{R}}\;\right)\right].\label{Chapitre SSS muSR}
\eea
Comparing the zero-th order terms, given by Eq. (\ref{SCHWA}) and Eq. (\ref{SCAbis}), and the first order ones, given by Eq. (\ref{Chapitre SSS lSR}) and Eq. (\ref{Chapitre SSS muSR}), we reach the conclusion that the expansion in powers of $m^{2}$ is only valid for $R\lesssim R_{V}\equiv\left(m^{-4}R_{S}\right)^{1/5}$. Hence, the functions (\ref{Chapitre SSS lSR}) give the first non trivial corrections to the Schwarzschild solution for distances below the Vainshtein radius. In section \ref{section numerics}, it will be seen they agree very well with the numerical results. Notice also that those corrections are non analytic in the expansion parameter $\epsilon$ of the perturbation theory of section \ref{PERTVAC}.
It is also interesting to notice that the expansion presented above and the ``small distance'' perturbative expansion of subsection \ref{SDExp} both yield the same ``Vainshtein radius'' $R_V$ as a limiting distance of their respective domain of validity.

One may wonder which of the expansions (\ref{expansion in a}) and (\ref{expansion in m}) is the most appropriate for describing the solution of the full NLPF theory. There is no simple answer to this question since, eventually, all the orders are needed to fully recover the solution. Still, a simple but instructive calculation can be done: one can check that for $R\lesssim (R_{V}m)^{\frac{8}{7}}R_{V}$, the correction to the Newtonian potential $\nu_{N}\sim -R_{S}/R$ due to the GR nonlinearities (\emph{i.e} the first correction to $\nu_{N}$ in $\nu_{0}^{(m)}$) dominates over the correction due to the DL nonlinearities (\emph{i.e} the first correction to $\nu_{N}$ in $a^{4}\nu_{0}^{(a)}$). In mathematical terms, this reads $(R_{S}/R)^{2}\gtrsim\left(m R \right)^{2} \sqrt{{R_{S}}/{R}}$. In particular, this means that for sources of radius $R_{\odot}\ll  (R_{V}m)^{8/7}R_{V}$ (which is usually the case for sources we considered in our numerical investigations), the GR solution is a better approximation inside the source than the DL solution. We were able to confirm this simple analysis numerically (\emph{cf.} Fig. \ref{fig nu comparison GR}). 

Note that here, for the sake of simplicity, we have followed \cite{Damour:2002gp} and discarded any integration 
constant appearing in the process of obtaining the successive terms (beyond the GR part (\ref{SCHWA})) in the expansion (\ref{expansion in m}). Such constants appear in particular at each order via a double integration of the Bianchi identity. If there is however a solution with such an expansion that can extend from $R=0$ to $R=\infty$, there is no reasons for such constants to vanish, and they should be kept explicit until one does the matching with the source.

\subsection{Expansion inside the source}
\label{INSIDE}
Let's now turn to the interior of the source, that we assume to be of constant density $\rho$ and of radius $R_{\odot}$.
In order to gain some understanding of the solution near the origin, it is possible to expand the solution as
\be\label{expansion in x}
\mu=\sum_{n=0}^{\infty}\mu^{(x)}_{n}\; x^{2n},\quad \lambda=\sum_{n=1}^{\infty}\lambda^{(x)}_{k}\; x^{2n},\quad \nu=\sum_{n=0}^{\infty}\nu^{(x)}_{n}\; x^{2n},\quad P=\sum_{n=0}^{\infty}P^{(x)}_{n}\; x^{2n},
\ee
where we have defined the rescaled variable 
\be
x\equiv \frac{R}{R_{S}}.\nonumber
\ee
Note that the expansion of the function $\lambda$ starts at $n=1$ (in agreement with condition (\ref{condition on lambda at the origin})). These series depend on 3 integration constants, that can be chosen to be $\mu^{(x)}_{0}\equiv\mu(R=0)$, $\nu^{(x)}_{0}\equiv\nu(R=0)$ and  $P^{(x)}_{0}\equiv P(R=0)$.
The other coefficients $\mu^{(x)}_{n>0},\lambda^{(x)}_{n>0},\nu^{(x)}_{n>0}$ are then functions of these integration constants $\mu^{(x)}_{0},\nu^{(x)}_{0},P^{(x)}_{0}$ and the mass $m$ of the graviton. These coefficients can be found solving order by order in $x^{2}$; a few of them are given for the AGS potential in Appendix \ref{series inside the source}.

\section{Weak-field approximation}
\label{WFA}
As we have just seen, different expansion schemes can be used in order to get understanding of the putative solution. However, those schemes all fail to describe correctly all the key features of the solution. More specifically, the DL scheme (section \ref{subsection results of DL}) describes well the Vainshtein crossover, but misses the Yukawa decay. The perturbation theory approach (section \ref{PERTVAC}) catches the Yukawa decay, but misses the Vainshtein crossover, while the expansion around GR misses both the Yukawa decay and the Vainshtein crossover (section \ref{EXPINV}).  The purpose of this section is to introduce an expansion scheme, called here weak-field approximation, that will retain both the Yukawa decay and the Vainshtein crossover.

We expand the full system of Einstein equations together with Bianchi identity in 
powers of $\lambda$,  $\nu$,  $\mu$, and assume the smallness of each function $\mu$, $\nu$ and $\lambda$ as well as of  their derivatives:
\be
\{\lambda, \nu, \mu\}\ll 1,\quad \{R\lambda', R\nu', R\mu'\}\ll 1,\quad R^2\mu''\ll 1.\label{assumption}
\ee
Expanding (\ref{EOMFULL}) and using our assumptions (\ref{assumption}) we obtain,
\bse
\label{expansion}
\begin{align}
\frac{\lambda'}{R}+\frac{\lambda}{R^2}  &=
	-\frac{m^2}{2}\left(\lambda+3\mu+R \mu' + \mathcal{O}(\nu^2) \right) + \frac{\rho}{M_P^2},\label{expansion1}\\
\frac{\nu'}{R}-\frac{\lambda}{R^2}&=\frac{m^2}{2}\left(\nu+2 \mu +\mathcal{O}(\lambda^2)+
	\mathcal{O}(R^2\mu'^2)\right),\label{expansion2}\\
\lambda -\frac{R\nu'}{2}&=
	-R^2\left(\frac{\mu'^2}{4}+\frac{2\mu\mu'}{R}+\frac{\mu\mu''}2\right)
	+\mathcal{O}(\nu^2)+\mathcal{O}(\lambda'^2R^2)+\mathcal{O}(\nu\lambda'R)\nonumber\\
		&\qquad +\mathcal{O}\left(\{\nu,\lambda'\}\times\{\mu,\mu'R,\mu''R^2\}\right).\label{expansion3}
\end{align}
\ese
Note that $P$ in the r.h.s. of (\ref{expansion2}) disappears as a consequence of the conservation equation (\ref{MATTCONS}) and the 
assumption of the weak field regime, similar to the DL case.
One can recognize, in the r.h.s. of Eq. (\ref{expansion3}),  the non-linear DL part responsible for the Vainshtein mechanism.
Note that so far we did not make any assumptions about relation between $\lambda$, $\nu$, $\mu$ or between 
functions and their derivatives, and this is the reason why we kept the quadratic terms in $\mu$ on r.h.s. of (\ref{expansion3}).

To further simplify the system of equations (\ref{expansion}) we assume that the quadratic terms $\sim\nu^2$ are small in comparison to the linear terms in r.h.s. of (\ref{expansion1}), i.e., 
\be
\nu^2 \ll \text{max}\left\{\lambda,\mu\right\},\nonumber
\ee
and similarly that the term $\sim\lambda^2$ in the r.h.s. of (\ref{expansion2}) is negligible compared with $\nu$. 
We also assume that the quadratic terms containing $\mu$ and its derivatives are negligible compared to the linear terms containing $\mu$.
It is important to note that the last assumptions eliminates the quadratic terms in Eqs.  (\ref{expansion1}) and (\ref{expansion2}), 
while we have to keep the quadratic terms in $\mu$ in (\ref{expansion3}),
since the linear term in $\mu$ is absent in this equation. 
Note, that if we assumed at this step that $\mu\sim\nu\sim\lambda$, then the quadratic terms in $\{\mu,\mu',\mu''\}$ in the r.h.s. of (\ref{expansion3}) 
would be dropped. 
It is important, however, that we do not assume any relations between $\{\mu,\mu',\mu'' \}$ and $\{ \nu,\lambda, \nu',\lambda' \}$, and thus we keep these terms. This will be the key feature of the expansion scheme introduced here.
As a result we obtain the system
\bse
\label{wfbis}
\begin{align}
\frac{\lambda'}{R}+\frac{\lambda}{R^2}  &= -\frac{m^2}{2}\left(\lambda+3\mu+R \mu' \right)+8\pi G_N\rho ,\label{wf1bis}\\
\frac{\nu'}{R}-\frac{\lambda}{R^2}&= \frac{m^2}{2}\left(\nu+2 \mu \right),\label{wf2bis}\\
\frac{\nu'}{2R}-\frac{\lambda}{R^2} &=- Q(\mu,\mu',\mu''),\label{wf3bis}
\end{align}
\ese
with $Q$ given by (\ref{Q}). 
It is then convenient to use once again the rescaled radial coordinate $z=Rm$ introduced in subsection \ref{PERTVAC}, such that we obtain from (\ref{wfbis}),
\bse
\label{wf}
\begin{align}
\frac{\lambda'}{z}+\frac{\lambda}{z^2}  &=
	-\frac12\left(\lambda+3\mu+\xi \mu' \right)+\bar{\rho},\label{wf1}\\
\frac{\nu'}{z}-\frac{\lambda}{z^2}&=\frac12\left(\nu+2 \mu \right),\label{wf2}\\
\frac{\lambda}{z^2}-\frac{\nu'}{2z} &= Q(\mu,\mu',\mu''),\label{wf3}
\end{align}
\ese
where 
\ba
Q(\mu,\mu',\mu'')=-\left(\frac{\mu'^2}{4}+\frac{\mu\mu''}{2}+\frac{2\mu\mu'}{z}\right),
\ea
and we introduced $\bar\rho\equiv 8\pi G_N\rho/m^2$.
This system of equation corresponds to the one left over in the limit considered here. As will be seen, it has the required properties.

First, note that it is possible to rewrite this system as a single ODE on $\mu$ and two algebraic relations between $\lambda, \nu$ and $\mu$ and its derivatives. Indeed,  
expressing $\lambda$ from (\ref{wf2}) in terms of $\nu$, $\nu'$ and $\mu$ and substituting it to Eqs.~(\ref{wf1}) and (\ref{wf3})
we find two equations containing only  $\nu$ and $\mu$ (and their derivatives):
\bse
\begin{align}
\nu''+\frac{2 \nu'}{z}-\frac{1}{4} \left(z^2+6\right) \nu-\frac{1}{2} z \mu'-\frac12 \left(z^2+3\right) \mu&=\bar\rho,\label{wfA1}\\
-\frac{\nu'}{2z}+\frac{\nu}{2}+\mu +Q&=0.\label{wfA2}
\end{align}
\ese
Taking the derivative of (\ref{wfA2}) with respect to $z$ and substituting the resolved $\nu''$ into (\ref{wfA1}) we obtain a system of two equations:
\bse
\label{wfB}
\begin{align}
\frac{z^2+3 }{z}\nu'-\frac14 \left(z^2+6\right)\nu -\frac{1}{2} \left(z^2+3\right)\mu +\frac{z}{2}\left(4Q' +3 \mu'\right)&=\bar\rho,\label{wfB1}\\
-\frac{\nu'}{2z}+\frac12\nu+\mu + Q&=0.\label{wfB2}
\end{align}
\ese
From the above system of equations (\ref{wfB}) we can find $\nu'$ in terms of $\mu$ and its derivatives
and  $\nu$ in terms of $\mu$ and its derivatives,
\bse
\begin{align}
\nu(z)&=\mathcal{F},\nonumber\\
\nu'(z)&=z \mathcal{F} + 2z (Q+\mu).\nonumber
\end{align}
\ese
where we denoted
\be
\mathcal{F} \equiv -\frac{2}{3\left(z^2+2\right)}
\left[z \left(4Q' +3\mu'\right)+ \left(z^2+3\right) \left(4Q+3\mu\right)-2 \bar\rho\right]
\ee
Thus we arrive to the following single ODE on $\mu$,
\be
\frac{d\mathcal{F}}{dz} = z \mathcal{F} + 2z (Q+\mu).\nonumber
\ee
The above equation can be  rewritten in a closed form,
\be
\label{master}
\begin{aligned}
\left(2+z^2\right)\left(4Q''+3\mu''\right)+\frac{2}{z}\left(4+z^2\right)\left(4Q'+3\mu'\right)
- \left(z^2+4\right)\left(4Q+3\mu\right)-Q\left(z^2+2\right)^2\\ 
=(2+z^2)\frac{2\bar\rho'}{z} - 2(4+z^2)\bar\rho.
\end{aligned}
\ee
Note that Eq.~(\ref{master}) is of the fourth order, meaning that the number of boundary conditions needed 
to solve the equation is four. This is in agreement with number of boundary conditions needed to specify 
for Eqs.~(\ref{wfbis}).
Eq.~(\ref{master}) can further be simplified under appropriate assumptions. First of all, let us recover the 
solution (\ref{linsolution}) for the linearized equations. To do so,  we drop in (\ref{master}) all non-linear terms, i.e. all 
the terms containing $Q$ and its derivatives, as well as the source terms (containing $\bar\rho$ and $\bar\rho'$) to get
\ba
\left(2+z^2\right)\mu''+\frac{2}{z}\left(4+z^2\right)\mu'-\left(z^2+4\right)\mu =0 \label{linearized}.\nonumber
\ea
The non-growing at the infinity solution of the above equation reads
\be
\mu=\frac{C e^{-z} \left(z^2+z+1\right)}{z^3},\nonumber
\ee
which coincides with Eq. (\ref{linsolution}) and catches the Yukawa decay.

Another limit can be obtained assuming in (\ref{master}) as follows. 
We assume $Qz\ll Q'$ and $\mu z\ll \mu'$ together with $z\ll 1$.
Neglecting sub-leading terms in (\ref{master}), one obtains,
\ba
\left(4Q''+3\mu''\right)+\frac{4}{z}\left(4Q'+3\mu'\right)
=\frac{2\bar\rho'}{z} -4\bar\rho.\label{smallz}\nonumber
\ea
The above equation can be integrated twice to give,
\be
2Q+\frac{3}{2}\mu=\frac{1}{z^3}\left(C_0+\int_0^z z^2\bar\rho dz\right)+C_1. \label{DLeq}
\ee
where in the last expression we also assumed,
$$\int_0^z z^2\left(\int_0^z \bar\rho \tilde{z}d\tilde{z} \right)dz \ll \int_0^z z^2\bar\rho dz$$
The constant of integration $C_1$ in (\ref{DLeq}) must be set to $0$ to get correct behaviour of $\mu$ at infinity
(otherwise the solution is non-flat at large distances). 
Similarly, $C_0=0$, by matching $\mu$ inside the Vainshtein radius and outside the source.
Restoring physical units, $R=z/m$, one can see that we arrived to the previously found DL equation (\ref{mueq}) catching the Vainshtein crossover.
Furthermore, the analytic solution of the weak field system (\ref{wf}) will be seen to reproduce very well the numerical integration (in the expected range) in the following section.

\section{Integration of the full system: Numerical results}

\label{section numerics}

We have integrated the equations of motion (\ref{EOMFULL}) and (\ref{MATTCONS}) using two different approaches. The first one is based on a resolution by relaxation, which turns out to be very well adapted to the problem. This method and the related results are presented respectively in the subsections \ref{subsection Relaxation method} and \ref{subsection numerical results}. The second approach, which is described in the subsection \ref{subsection Shooting method} consist in a shooting method {\it\`a la} Runge-Kutta. Both methods perfectly agree, demonstrating the robustness of our numerical investigations.
A reader mainly interested in the numerical results can  read only subsection \ref{subsection numerical results} which is self-contained.

\subsection{Relaxation approach: method and boundary conditions}\label{subsection Relaxation method}
The basic idea of the relaxation method consists in choosing an initial configuration for the solution, and then deforming it until the equation we want to solve is satisfied with a good enough precision. 
The initial guess is of prime importance, since it will determine whether the iteration converges or not. In the case of the NLPF equations of motion (\ref{EOMFULL}) and (\ref{MATTCONS}), a clever guess is needed: indeed, starting from a naive guess (\emph{e.g} all the functions to zero) does not lead to a converging iteration. Here is where the DL plays a crucial r\^ole in finding a solution of the full system: we can use the DL solution as a guess for the full solution. Note that the DL being simpler than the full system, the DL solution itself is not too hard to find (no specific guess is needed).

The advantage of the relaxation approach is that boundary conditions can be easily enforced. In particular, this method allows to fix boundary conditions at different points, for example at the two ends of the interval on which we want to solve the equations. This flexibility in the boundary conditions fixing makes this method very well fitted for static solution studies in which we want to impose conditions both in the center and very far from the source.

In our case, we worked with the dimensionless variable 
\be
\xi\equiv R/R_{V}, \nonumber
\ee
as well as with the rescaled functions $w$, $u$, and $v$ defined by 
\bea
w &=& a^{-2} \mu, \nonumber\\
v &=& a^{-4} \nu, \nonumber\\
u &=& a^{-4} \lambda. \nonumber
\eea
We fixed the boundary condition at $\xi\equiv R/R_{V}=0$ and at some distance $\xi_{\infty}$ which is chosen to be much larger than the Vainshtein radius (typically $\xi_{\infty}\sim 100$).
The boundary conditions in the DL case are given by the DL linear regime (\ref{DLsolinf}) and read
\bea
\dot{w}(\xi=0)&=&0,\nonumber\\
u(\xi=0)&=&0,\nonumber\\
\dot{w}(\xi=\xi_{\infty})&=&-\frac{2}{\xi_{\infty}^{3}},\nonumber\\
v(\xi=\xi_{\infty})&=&-\frac{4}{3\;\xi_{\infty}},\nonumber\\
P(\xi=\xi_{\infty})&=&0.\nonumber
\eea
One may notice that the condition on $w$ at infinity is imposed on $\dot{w}$ rather than on $w$. This is not necessary, but we found it easier to proceed that way in practice. Once the DL solution has been found, we can turn to the fully nonlinear system. In that case, the boundary conditions are given by the linear solution (\ref{linsolution}) and read 
\bea
\dot{w}(\xi=0)&=&0,\nonumber\\
u(\xi=0)&=&0,\nonumber\\
{w}(\xi=\xi_{\infty})&=&\mathcal{C}\times\frac{2\;a^{2}}{3\;\xi_{\infty}}\left(1+\frac{1}{a\xi_{\infty}}+\frac{1}{(a\xi_{\infty})^{2}}\right)e^{-a\xi_{\infty}},\nonumber\\
v(\xi=\xi_{\infty})&=&-\mathcal{C}\times\frac{4}{3}\frac{e^{-a\xi_{\infty}}}{\xi_{\infty}},\nonumber\\
P(\xi=\xi_{\infty})&=&0.\nonumber
\eea
We have explicitly kept in these boundaries conditions the integration constant $\mathcal{C}$ defined in Eq. (\ref{linsolution})  (while $a$ is defined as in Eq. (\ref{def a})). This procedure allows to check numerically that the choice we made $\mathcal{C}=1$ in Eq. (\ref{def C}) is in agreement with numerics.

In our study of  NLPF, we have implemented the relaxation method described in \cite{NumericalRec}, using the \textsf{C++} language. We typically used a grid of $N\sim 5\times 10^{4}$ points and aimed at a precision of $10^{-6}$. We always checked that our numerical solutions were stable under the change of the step size of the grid and the precision.

The density of the star is assumed to be given by $\rho=M/(4\pi R_{\odot}^{3})$ inside the star, and null outside.

\subsection{Numerical results}\label{subsection numerical results}

To summarize here shortly our main results, we were able to show that one can obtain a non singular solution showing the recovery of GR {\it \`a la} Vainshtein. The relaxation method allowed us to solve the equations of motion (\ref{EOMFULL}) and (\ref{MATTCONS}) on a wide range of the parameter $a=R_{V}m$ ($a \in [10^{-3},0.6]$) and for very large intervals (typically for $R$ between $0$ and $\sim 100 \; R_{V}$).
In the following we first discuss the behaviour of the functions $\lambda$ and $\nu$ which appear in the ``physical'' metric $g_{\mu \nu}$ (subsection \ref{LaNu}), comparing their behaviour in NLPF and GR, and showing in particular how our numerical integration agrees with the corrections to GR as predicted by Vainshtein. Then  (subsection \ref{PGR}) we turn to discuss the obtained behaviour of the pressure inside the star, comparing again the numerical solution with the one obtained in GR (recall that the star's density is assumed to be constant).  Last we discuss our numerical result for the function $\mu$ appearing in the ``non physical'' metric with our gauge choice (\ref{lammunu}).

\subsubsection{Behaviour of the functions  $\lambda$ and $\nu$}
\label{LaNu}
\begin{figure}[h]
  \begin{center}
    \includegraphics[width=0.5\textwidth]{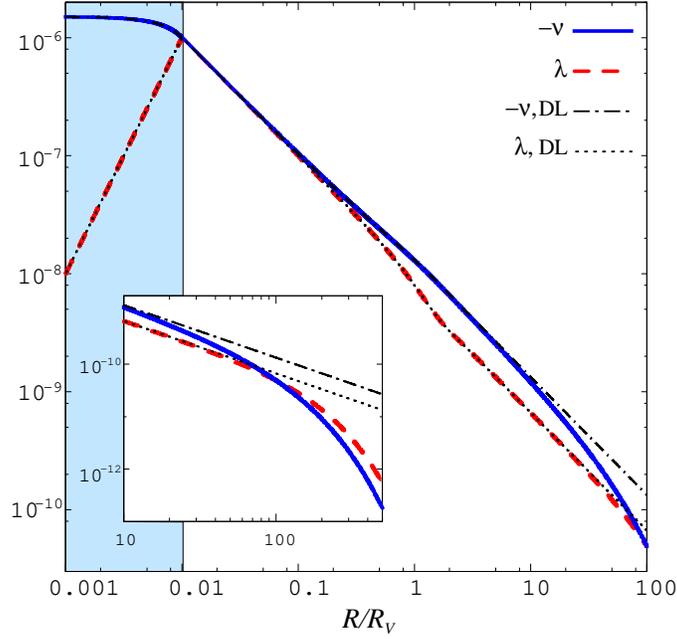}
    \caption{ \label{fig1} Plot of the metric functions $-\nu$ and $\lambda$ vs. $R/R_V$, in the full nonlinear system and the decoupling limit (DL), with a star of radius $R_{\odot} =10^{-2} R_V$
and $ m \times R_V = 10^{-2}$. For $R \ll R_V$, the numerical solution is close to the GR solution (where in particular $\nu \sim -\lambda \sim- R_S/R$ for $R > R_{\odot}$).
For $R \gg R_V$, the solution enters a linear regime. Between $R_V$ and $m^{-1}$, where the DL is still a good approximation, one has $\nu \sim - 2 \lambda\sim -4/3 \times R_S/R$. At distances larger than $m^{-1}$ the metric functions decay {\it \`a la} Yukawa as appearing more clearly in the insert. The latter shows the same solution but for larger values of $R/R_V$, and in the range of distance plotted there, the numerical solutions are indistinguishable from the analytic solutions of the linearized field equations Eqs.~(\ref{linsolution}). This plot was already presented as the Fig.1 of  Ref. \cite{us}.}
  \end{center}
\end{figure}

The Fig. \ref{fig1} summarizes our main results and illustrates the following points: 
\begin{itemize}
\item at short distances  ($R\ll R_{V}$), the solution is very close to the GR solution, and in particular $\nu \sim -\lambda \sim- R_S/R$ in the neighborhood of the source,
\item at large distances ($R\gg R_{V}$), the solution is in the linear regime (\ref{linsolution}). Between  $R_{V}$ and $m^{-1}$, the solution is very close to the DL solution  (\ref{table behavior vainshtein solution}), and in particular $\nu \sim - 2 \lambda\sim -4/3 \times R_S/R$. Beyond $m^{-1}$, the solution is exponentially decreasing,
\item the behaviour at infinity allows to check that the choice of integration constant we made in Eq. (\ref{def C}) is consistent with numerics. More precisely, we tried to solve numerically the equations of motion with a integration constant $\mathcal{C}$ slightly different from 1, and we noted that the relaxation method was not converging, unless $|\mathcal{C}-1|\ll 1$.
\end{itemize}

\begin{figure}[h]
  \begin{center}
    \includegraphics[width=0.5\textwidth]{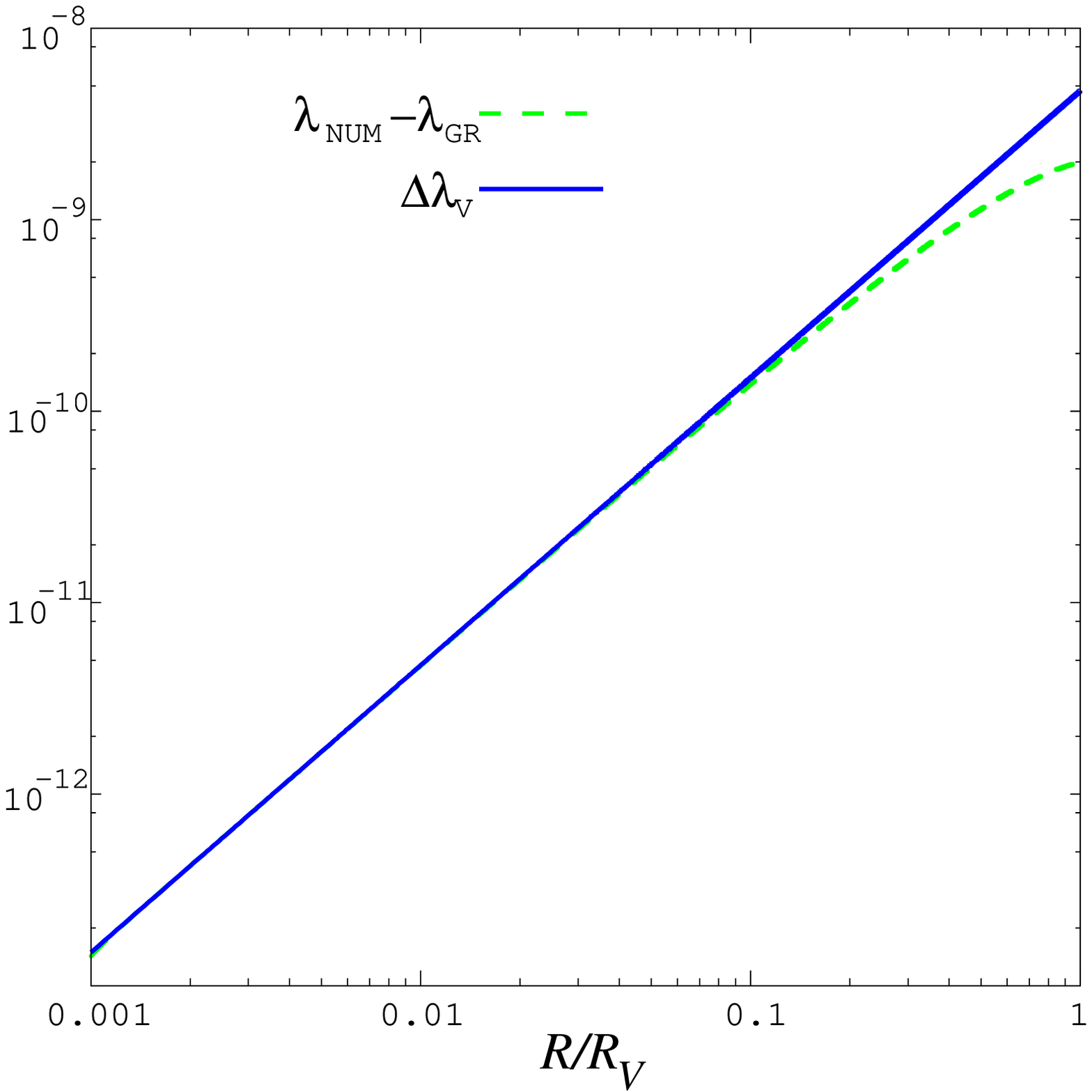}
    \caption{\label{fig vainshtein-correction}  Plot of the difference between the numerical solution for $\lambda$, $\lambda_{NUM}$ and the one of GR, $\lambda_{GR}$. This is compared to the computation of the same quantity using the first term in the expansion in $m^2$ proposed by Vainshtein, $|\Delta \lambda_{V}|=\frac{\sqrt{2}}{3}\left(m R \right)^{2} \sqrt{\frac{R_{S}}{R}}. $ Both functions are plotted  \emph{vs.} $R/R_V$,  for a source of radius $R_{\odot} =10^{-3} R_V$ and a choice of parameters such that  $a\equiv m \times R_V = 10^{-2}$. It can be seen that the numerical solution agrees very well   with the Vainshtein's expression given by Eq. (\ref{Chapitre SSS lSR}) for $R \ll R_V$. For $R \sim R_V$, the expansion in $m^2$ is, as expected, no longer a good approximation of the solution.}
  \end{center}
\end{figure}

We have seen in subsection \ref{subsection Expansion inside the Vainshtein radius} that the leading behaviour of the difference between the functions 
$\lambda$ and $\nu$ and their GR equivalent, $\lambda_{GR}$ and $\nu_{GR}$ is given by the expressions  (\ref{Chapitre SSS lSR}). It is possible to check that this theoretical prediction is very well satisfied by our numerical solution, as can be seen in 
Fig. \ref{fig vainshtein-correction}.  In this figure, we plotted the difference between the function  $\lambda$ obtained from our numerical integration and the one of GR, and compared this with the first correction to GR as computed using the Vainshtein mechanism. As expected, the two agree very well for distances smaller than the Vainshtein radius.

\begin{figure}[h]
  \begin{center}
    \includegraphics[width=.5\textwidth]{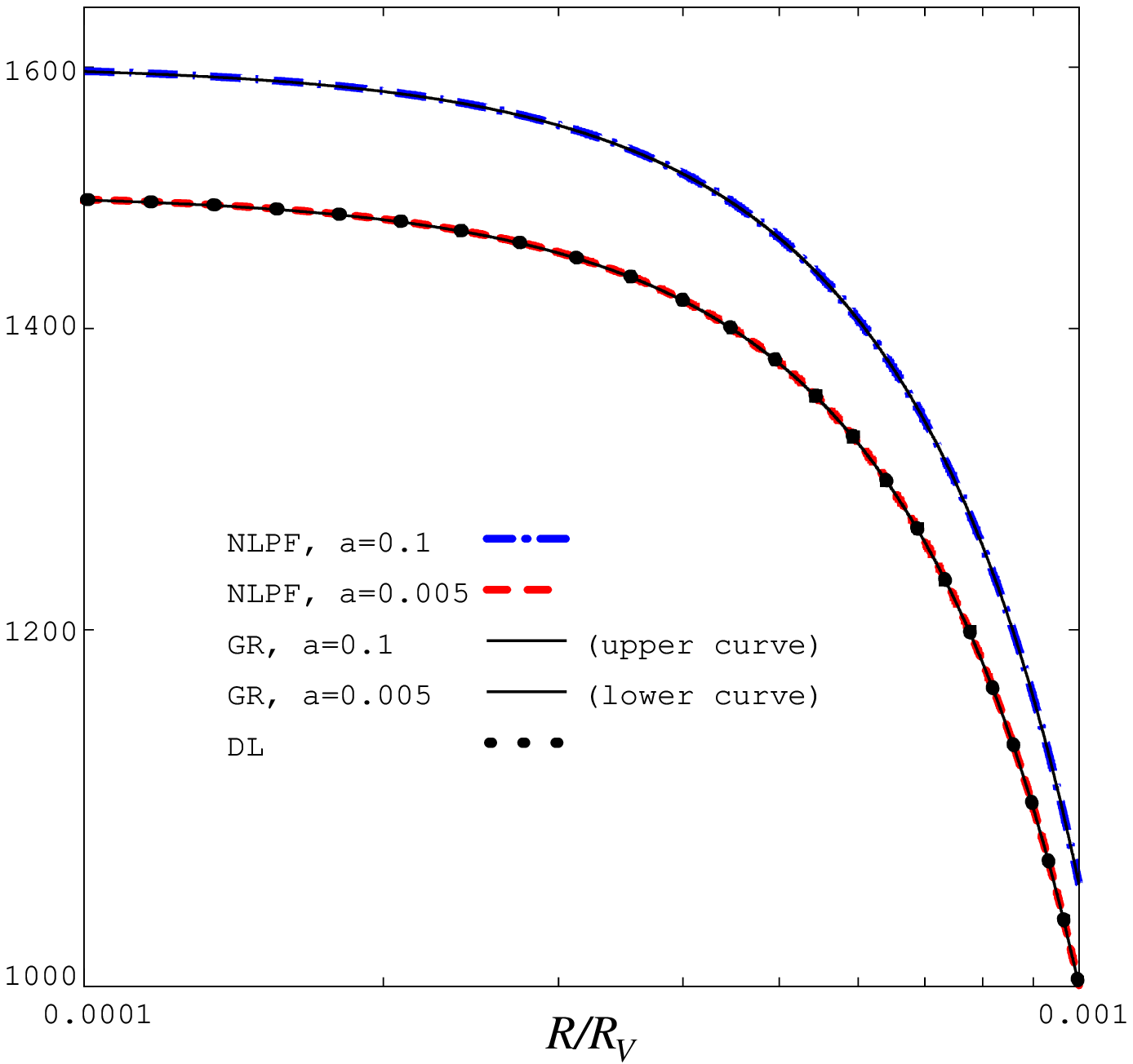}
    \caption{ \label{fig nu comparison GR} 
Plot of $-\nu \times a^{-4}$ vs. $R/R_V$ (the $a^{-4}$ factor is included for convenience such that, in the decoupling limit (DL), all plotted theories would exactly coincide) inside the source of radius $R_{\odot} = 10^{-3} R_V$, for three different values of $a \equiv  m \times R_V$. Along with the numerical solution, the GR analytical solution  for $\nu$ is plotted for each value of $a$. It can be seen that the numerical solutions perfectly agree with the GR behaviour, while slightly differing form the DL expression for large $a$. This illustrate the fact that inside the source, the nonlinearities coming from the GR terms of Eq. (\ref{EOMFULL}), and which are not taken into account in the DL, are important. For small $a$, both the GR and the numerical solutions agree with the DL solution, which encodes for most of the physics, in agreement with the fact that the DL corresponds to the limit $a\to 0$.}
  \end{center}
\end{figure}

Besides, we have shown at the end of the subsection \ref{subsection Expansion inside the Vainshtein radius} that deep inside the Vainshtein radius, the NLPF solution should be closer to the GR solution than to the DL solution, illustrating the fact that the GR nonlinearities are starting to play an important r\^ole. This is seen in Fig. \ref{fig nu comparison GR}, where 
we compared the function $\nu$ (after an appropriately chosen rescaling) for various values of the parameters. It appears on this figure that the numerical solutions  well agree with the GR behaviour, while slightly differing form the DL expression for large $a$. This illustrate the fact that inside the source, the nonlinearities coming from the GR terms of Eq. (\ref{EOMFULL}), and which are not taken into account in the DL, are important.

\subsubsection{Behaviour of the pressure $P$}
\label{PGR}
\begin{figure}[h]
 \begin{center}
   \includegraphics[width=0.5\textwidth]{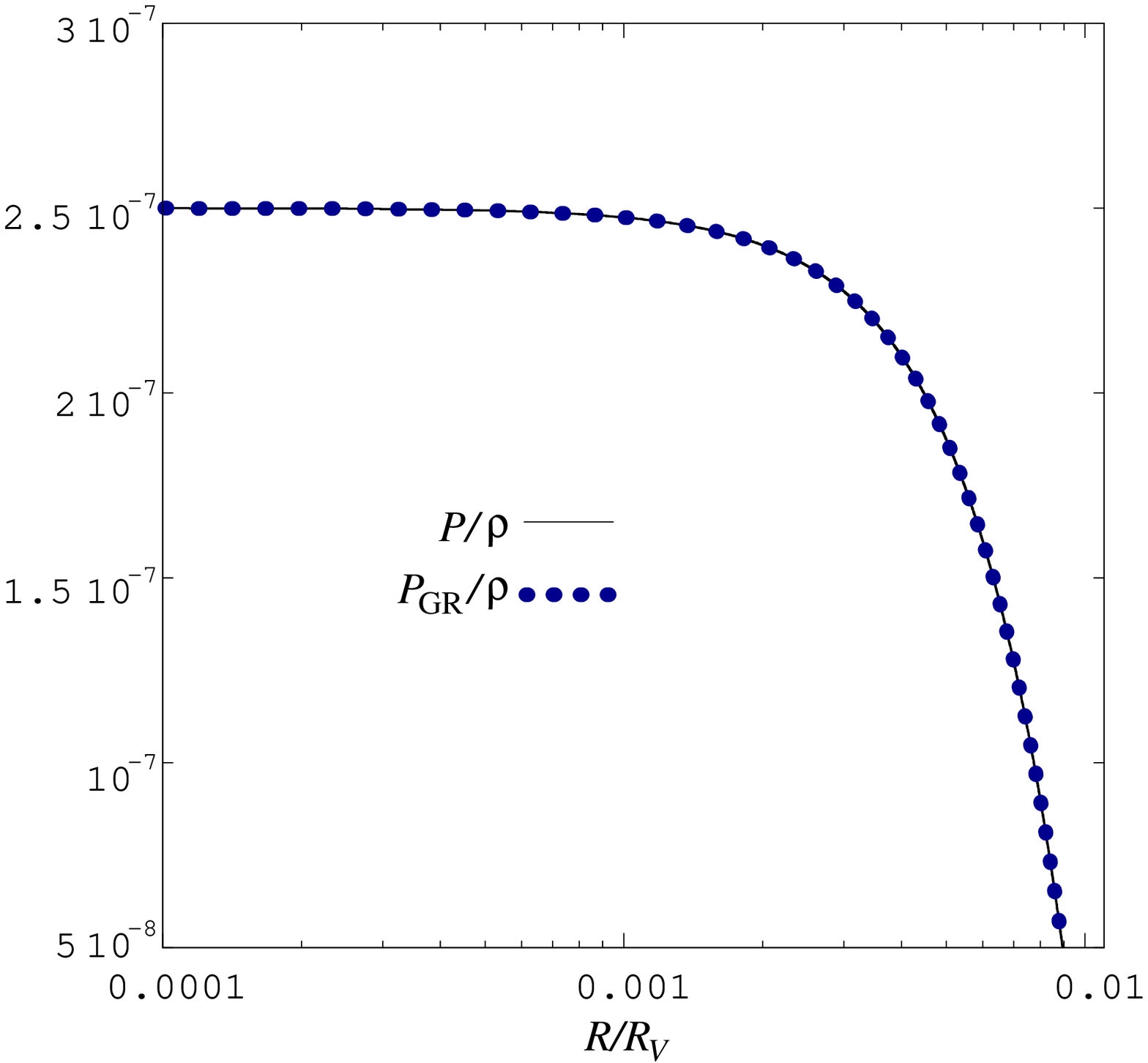}
    \caption{\label{fig pression} Plot of the functions  $P/\rho$  and $P_{GR}/\rho$ \emph{vs.} $R/R_V$,  for a source of radius  $R_{\odot} =10^{-2} R_V$, of pressure $P$, and of constant density $\rho$, and for a parameters choice such that $a\equiv m \times R_V = 10^{-2}$. The function  $P/\rho$ if found numerically, while  $P_{GR}/\rho$ is  the pressure computed in GR and is given by the analytical expression (\ref{pression gr}). The two curves are indistinguishable from each other, illustrating the fact that the Vainshtein's conjecture is valid inside the star.}
  \end{center}
\end{figure}

In GR, the pressure inside a static spherically symmetric source of constant density can be computed analytically to be
\be
P_{GR}=\rho \frac{\sqrt{1-\frac{R_S}{ R_{\odot}^3}R^2}-\sqrt{1-\frac{R_S }{ 
R_{\odot}}}}{3\sqrt{1-\frac{R_S}{ R_{\odot}}}-\sqrt{1-\frac{R_S}{ R_{\odot}^3}R^2}}\;.\label{pression gr}
\ee
This is compared on figure \ref{fig pression} with our numerical solution, confirming once again the validity of the Vainshtein's conjecture.

\subsubsection{The function $\mu$}
\label{MURESNUM}
\begin{figure}[h]
  \begin{center}
    \includegraphics[width=0.5\textwidth]{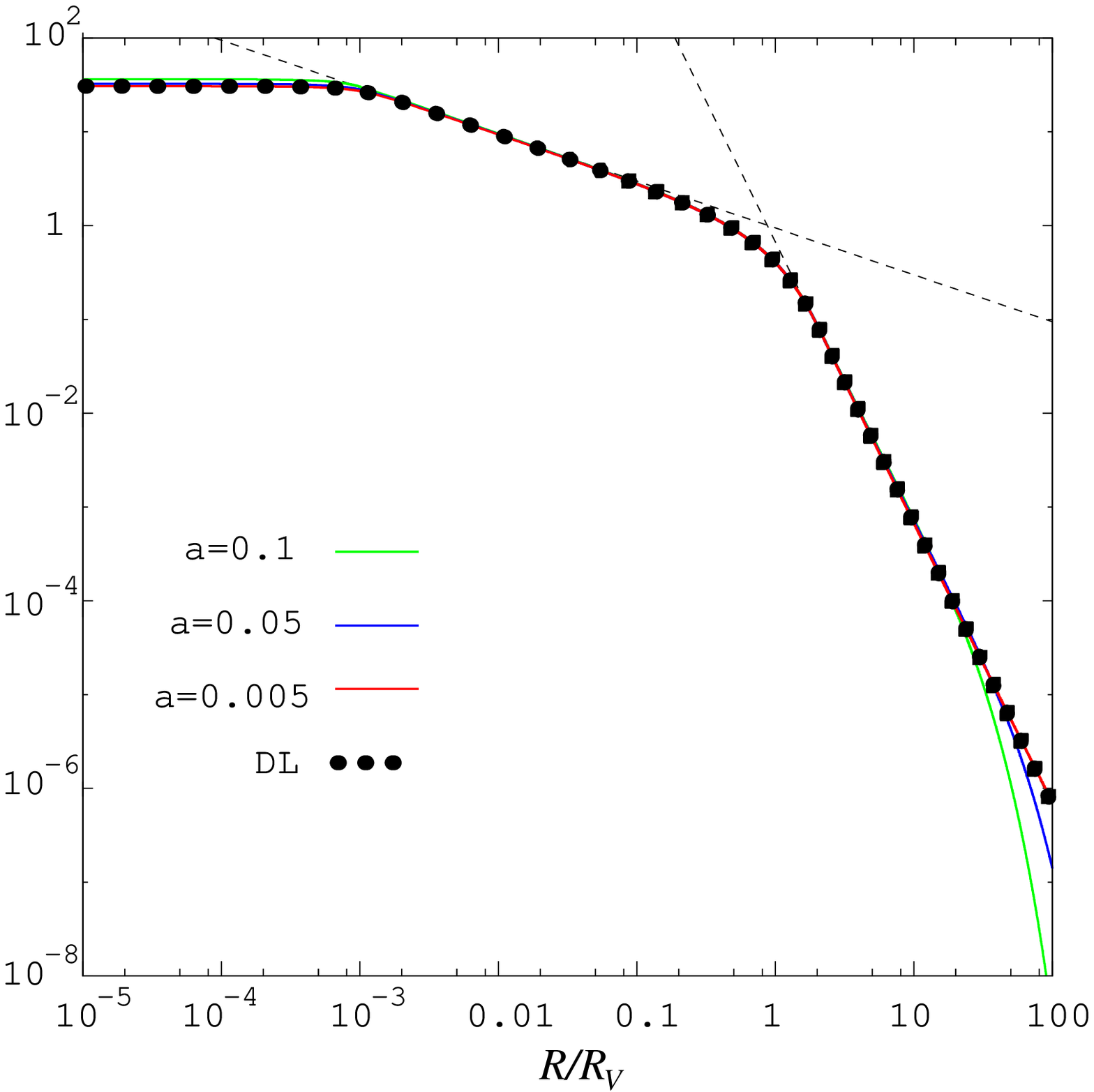}
    \caption{ \label{fig mu solution} 
Plot of $a^{-2}\mu$ vs. $R/R_V$ (the $a^{-2}$ factor is included for convenience such that, in the decoupling limit (DL), all plotted theories would exactly coincide) for a source of radius $R_{\odot} = 10^{-3} R_V$ for  three different values of $a \equiv  m \times R_V= 0.005,0.05,0.1$ as well as in the DL case (which corresponds to  $a\rightarrow 0$). The three DL regimes of Eq.  (\ref{table behavior vainshtein solution}) are clearly distinguishable: for $R<R_{\odot}$, $\mu\sim \text{const}$; for $R_{\odot}<R\ll R_{V}$, $\mu\sim\sqrt{(8R_{V})/(9R)}$; for $R_{V}\ll R\ll m^{-1}$, $\mu\sim2R_{V}^{3}/(3R^{3})$. For $R\gg m^{-1}$, we observe the Yukawa decay of the linear solution  (\ref{linsolution}).
At the resolution of the picture, the thee numerical solutions are hard to distinguish for distances $R$ smaller than $m^{-1}$. The next figure shows a zoom of this figure at small distances (namely inside the source) where the differences between the three solutions are easily seen}
  \end{center}
\end{figure}

\begin{figure}[h]
  \begin{center}
    \includegraphics[width=0.5\textwidth]{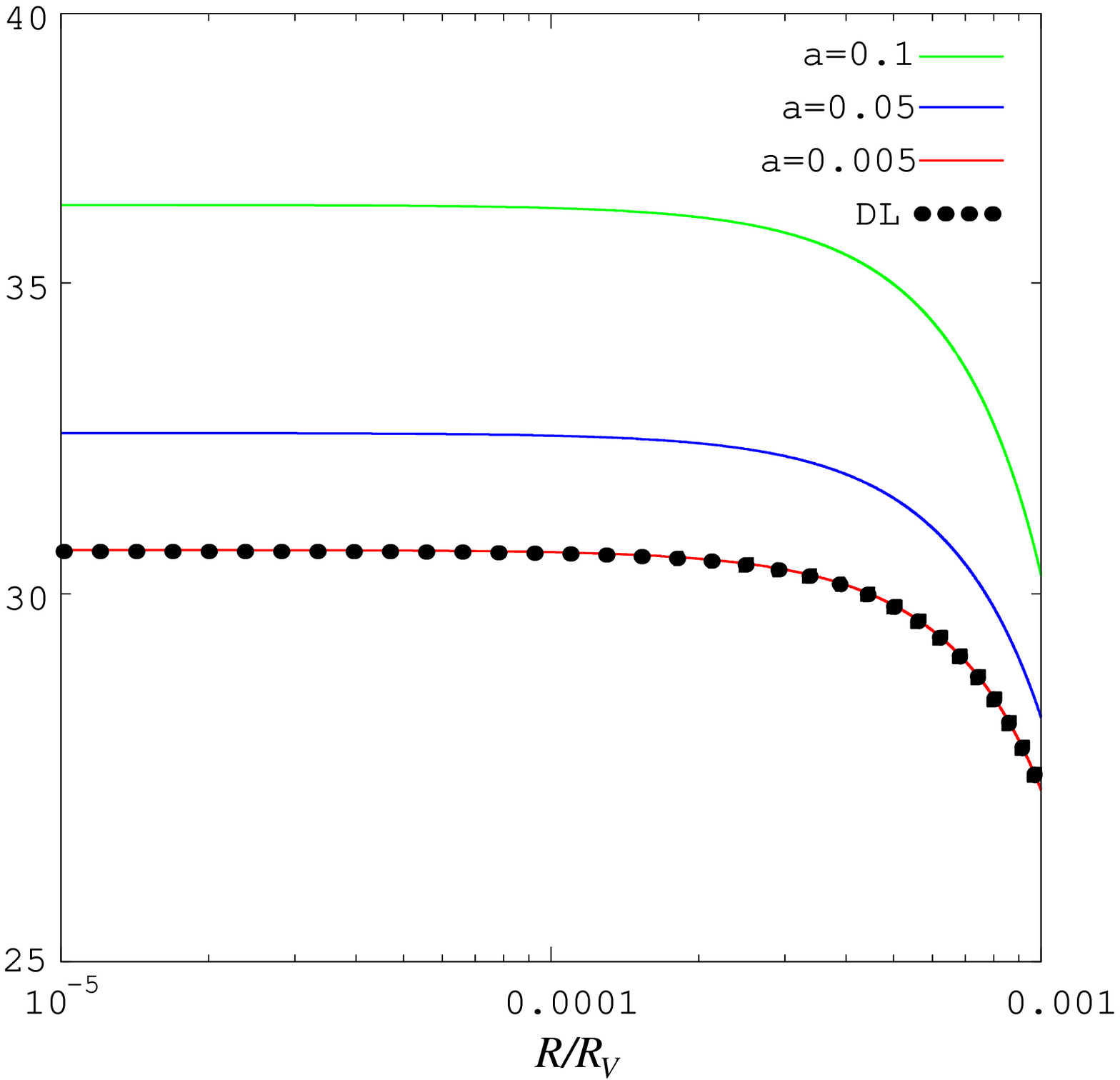}
    \caption{\label{fig mu solution zoom} Plot of $a^{-2}\mu$ vs. $R/R_V$ (the $a^{-2}$ factor is included for convenience such that, in the decoupling limit (DL), all plotted theories would exactly coincide) for a source of radius $R_{\odot} = 10^{-3} R_V$ for  three different values of $a \equiv  m \times R_V= 0.005,0.05,0.1$ as well as in the DL case. The picture  is a zoom of  Fig. \ref{fig mu solution}, inside the star. One can check that even if the three curves for $a = 0.005,0.05,0.1$ are close to each other and to the DL solution, they differ inside the star due to non-linearities not captured by the DL. One note that the full solution goes towards the DL solution when  $a\equiv R_{V}\times m$ tends towards $0$.  }
  \end{center}
\end{figure}

For the range of parameters investigated here, the numerical behaviour of the  function $\mu$ between  $R=0$ and $R=m^{-1}$ is very close to the  Vainshtein solution (\ref{table behavior vainshtein solution}) of the DL. Beyond  $m^{-1}$, the  Yukawa decay seen in the linear solution  (\ref{linsolution}) is found to occur. These various regimes are shown in Fig. \ref{fig mu solution} which plots the (rescaled) function $\mu$ for three different values of the parameters.
Figure \ref{fig mu solution zoom} shows the same, zoomed inside the star.
 We notice there that the solution of the full theory goes towards DL solution as the parameter $a\equiv R_{V}\times m$ tends towards $0$ and we recover the fact that the corrections to the DL due to the nonlinearities which are not taken into account in the DL are of order 
$\mathcal{O}(a^{2})$ and goes to zero as $a\rightarrow 0$.
This plot also shows that our numerical integration reproduces nonlinearities not present in the DL.

\subsection{Shooting method}\label{subsection Shooting method}
In addition to the relaxation approach that we presented in the previous subsections, we developed a shooting approach to solve the system of equations (\ref{EOMFULL}) and (\ref{MATTCONS}). This shooting approach, which is based on a direct Runge-Kutta integration, appears to be quite difficult to implement for the theory (\ref{S3}). Indeed, there appear numerical instabilities for distances beyond the Vainshtein radius and probably related to the growing mode (\ref{finalsol growing}). As a consequence of these instabilities, the initial conditions need to be extremely finely 
tuned\footnote{Note however that the shooting integration appears more stable 
for some NLPF theories which have a different interaction term from (\ref{S3}). This is the case for example for the BD theory (using the terminology of Ref. \cite{us}). In this theory, there is no admissible Vainshtein scaling, but only Q-scalings in the DL. However, in the DL and in the full non linear case, the shooting method appears much more stable than in the model considered here, even though the Q-scaling solution does not seem to continue into a non singular solution in the beyond DL regime.}. Nevertheless, we have been able to integrate the system of equations outward from a point deep inside the star up to a point further than the Vainshtein radius (typically $R\sim 3 R_{V}$). We have also been able to integrate inward starting from a point far from the source (\textit{e.g.}  $R\sim 3 R_{V}$) up to the core of the source. In all the integrations, we have made an extensive use of  series expansions of the solution in order to fix the initial conditions as well as possible. 

While integrating inwards from large distances (\emph{i.e.} $R\sim 3 R_{V}$), the solution adopts almost systematically a $Q$-scaling behaviour, similar to the one of Eq. (\ref{DLsolQ}), except if the initial conditions are extremely fine tuned to get the Vainshtein solution. In the cases in which the $Q$-scaling is found, our numerical solutions always encountered a singularity around the radius we identified in Eq. (\ref{radius singularity}) and which corresponds to the scale at which nonlinearities non taken into account by the DL start to play a r\^ole. Of course, this does not mean that a $Q$-scaling solution cannot be found but we were not able to find any and it convinced us to focus on the Vainshtein solution.

\begin{figure}[h]
  \begin{center}
    \includegraphics[width=0.5\textwidth]{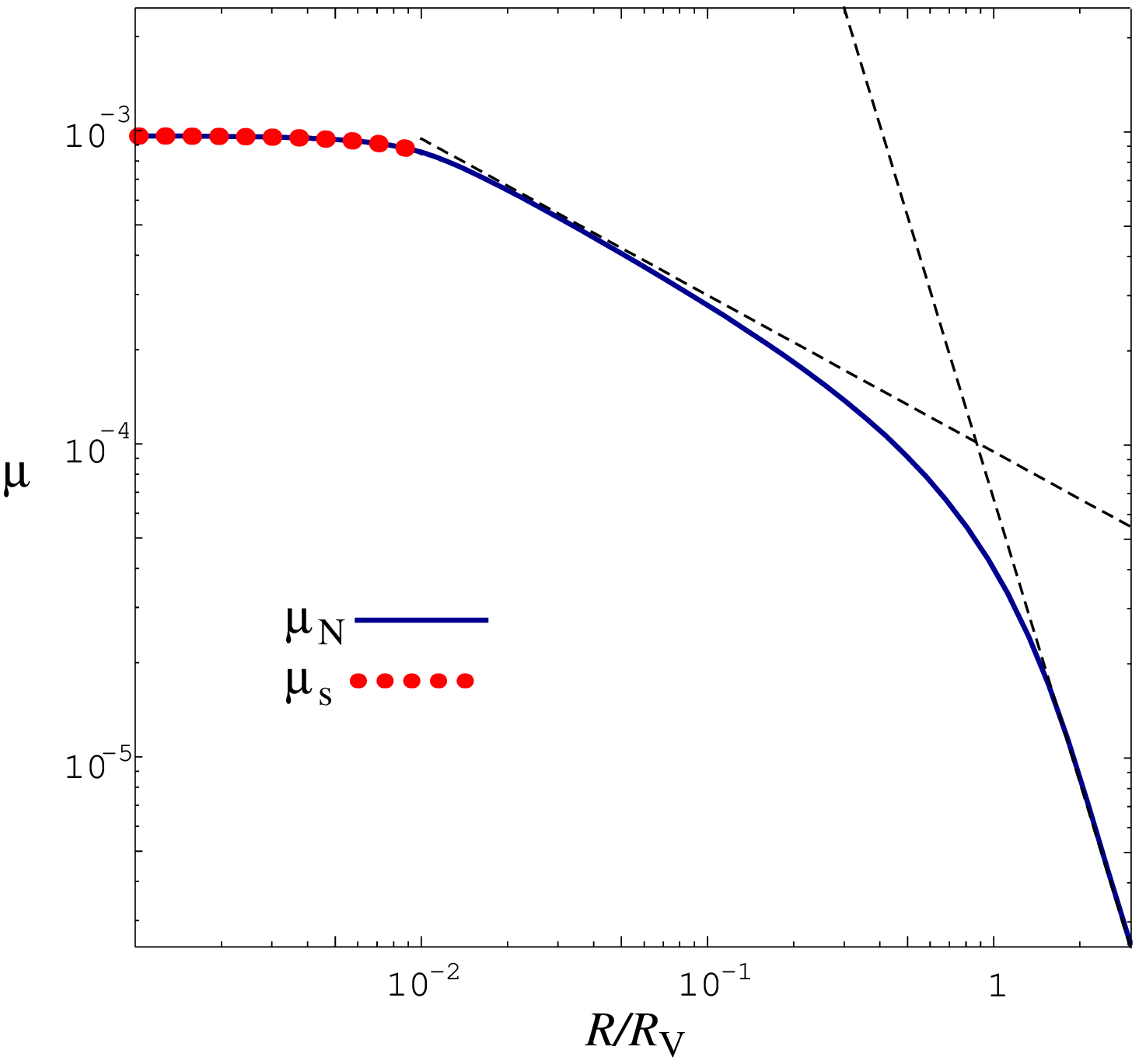}
    \caption{\label{fig shooting} Plot of the function $\mu$ ($\mu_N$ as given by the numerical integration) \emph{vs.} $R/R_V$, for a source of radius  $R_{\odot} =10^{-2} R_V$ and of constant density $\rho$, and for a choice of parameters such that $a\equiv m \times R_V = 10^{-2}$. The plot has been obtained using an direct integration {\it\`a la} Runge-Kutta. We also added the plot of series $\mu_S = \mu^{(x)}_{0}+\mu^{(x)}_{1}x^{2}+\mu^{(x)}_{2}x^{4}$, which was introduced in the main text as an expansion inside the star: the two curves are indistinguishable from each other.}
  \end{center}
\end{figure}

To obtain the latter, 
 we have found particularly convenient to integrate from a point located inside the source (typically $R_{min}=10^{-3}R_{\odot}$). To do so, we needed initial conditions at $R_{min}$, \emph{i.e.} $\mu(R_{min})$, $\mu'(R_{min})$, $\lambda(R_{min})$, $ \nu(R_{min})$ and $P(R_{min})$. These initial conditions are computed using the expansion (\ref{expansion in x}) where the three integration constants $\mu^{(x)}_{0}\equiv\mu(R=0)$, $\nu^{(x)}_{0}\equiv\nu(R=0)$ and  $P^{(x)}_{0}\equiv P(R=0)$ have been fixed using the values obtained through relaxation method.
 In order to be able to obtain a wide enough range of integration, the required relative precision on the  three integration constants $\mu^{(x)}_{0}$, $\nu^{(x)}_{0}$ and  $P^{(x)}_{0}$ is very stringent. For example,  if one wants to be able to integrate from $R_{min}=10^{-3}R_{\odot}$ to $R_{max}= 3 R_{V}$, in the case $a=R_{V}m=10^{-2}$ and $R_{\odot}=10^{-2}R_{V}$, the required relative precision on the integration constant $\mu^{(x)}_{0}$ is smaller than $10^{-8}$,
in order to ensure that the numerical solution at $R_{max}=3 R_{V}$ differs from the linear solution (\ref{linsolution}) (which is supposed to be an excellent approximation of the solution at that distance) by less than $5\%$. The result of such an integration is given in Fig. \ref{fig shooting}. This figure also shows part of the expansion (\ref{expansion in x}) inside the star. It agrees very well with the numerical solution.

 \subsection{Weak-field approximation}
\begin{figure}[h]
  \begin{center}
    \includegraphics[width=0.5\textwidth]{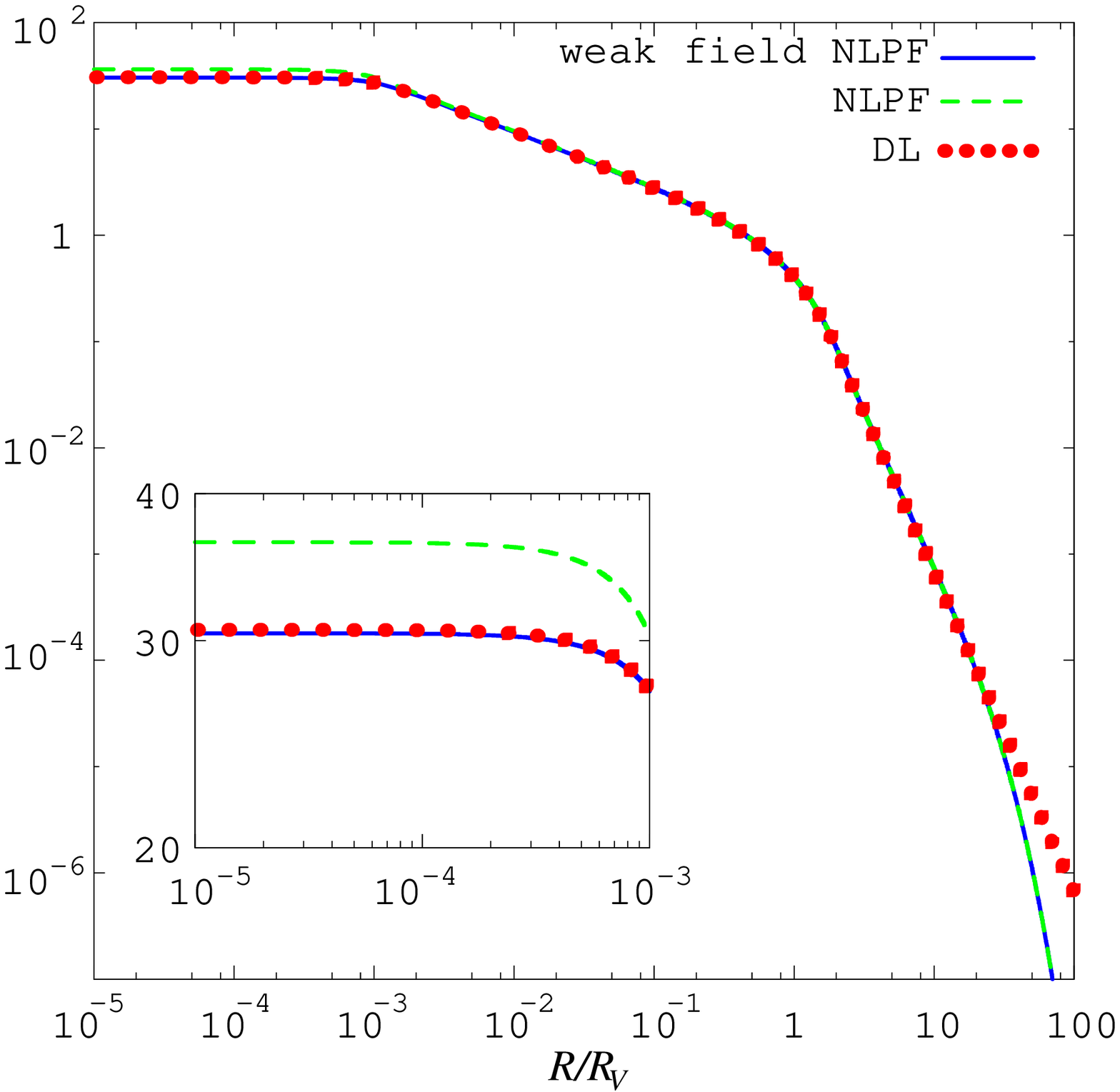}
    \caption{\label{figure mu-WF} Plot of the function $a^{-2}\mu$ (the $a^{-2}$ is added to make the comparison between $\mu$ and its DL behaviour easier), \emph{vs.} $R/R_V$,  for a source of radius $R_{\odot} =10^{-3} R_V$, of constant density $\rho$ and a choice of parameters such that $a\equiv m \times R_V = 10^{-1}$. The three curves corresponds respectively to the weak-field solution of the equation (\ref{master}), to the solution of the fully nonlinear theory and to the DL solution. One can see that the weak-field model allows to recover the  general behaviour of the full solution, except inside the star (as seen in the panel showing a zoom of the plots for small $R$) where it is very close to the DL, and  thus misses some nonlinearities.}
  \end{center}
\end{figure}

We have also been able to check numerically the weak field approximation introduced in section \ref{WFA}. Namely, as appears on Fig. \ref{figure mu-WF}, this approximation reproduces very well the numerical solution on a range of distances much larger than the decoupling limit, since this range extends at distances larger than  the graviton Compton length $m^{-1}$. At small distances, however, as also seen on figure \ref{figure mu-WF}, the week field approximation deviates from the numerical solution. This is expected since the week field approximation reduces there to the DL and does not capture some non linearities.

\section{Conclusions}
\label{CCL}
We have studied in detail the Vainshtein mechanism for Non Linear Pauli-Fierz theory as defined in section \ref{GENSSSS}. More specifically, we have been able to show in the case of a specific NLPF model (the one defined by the interaction term (\ref{S3})) that the recovery of the Schwarzschild solution  works as forecasted by Vainshtein. This proof was provided via numerical integration of the field equation using two different methods: a relaxation scheme and a shooting method. It also heavily used series expansions (to fix boundary/initial conditions) which were first presented  in this work. Our result, already presented in brief in \cite{usprl} disagrees with a previous work on the same issue \cite{Damour:2002gp}, which concluded that the Vainshtein mechanism was not working, because it could not find numerically any non singular asymptotically flat solution of the NLPF field equation. We also presented arguments, based on various expansions, showing that the solution at infinity is not uniquely defined by standard perturbation theory, but has non-perturbative hairs allowing to match sources at smaller distances. This, as well as the more sophisticated integration methods we used, could explain the differences between our results and the ones of Ref. \cite{Damour:2002gp}. Besides the model defined by the interaction term (\ref{S3}), we also investigated other models, in particular those which were shown in Ref. \cite{us} to possess only a non-Vainshtein like scaling (dubbed Q-scaling in Ref. \cite{us}) in the Decoupling Limit (DL). According to our numerical investigations, the DL solutions with Q-scaling, do not seem to continue into non singular solutions of the full field equations. We have also introduced a new limit, the weak-field approximation, which captures all the salient feature of the solution, including the Yukawa decay and the Vainshtein crossover. Of course, the theory considered is believed to suffer from various pathologies, like Boulware-Deser type of instabilities \cite{Boulware:1973my,Deffayet:2005ys,Creminelli:2005qk}. However, our investigations show for the first time that the Vainshtein mechanism can actually work in theories of massive gravity, some of which could be safe from the pathologies of NLPF theory. Several questions are however left unanswered. First, we have not been able to solve numerically the equation for dense objects (or even realistic stars). Indeed, when one increases the density of the object, the numerics becomes unstable and singularities are found to appear. There remains to understand if those singularities are physical or just numerical artifacts. Along the same line, it is not known if standard black holes can be recovered via the Vainshtein mechanism. Second, it is not clear if the solution we found is stable or not. The Boulware-Deser type of instability could show up if one would try to let the solution be time dependent or to be reached by a dynamical process such as spherical collapse. These and other issues are left for future investigations.

\begin{acknowledgments}
We thank T. Damour for interesting discussions and suggestions. The work of E.B. was supported by the TRR 33 ``The Dark Universe''.
 \end{acknowledgments}

\appendix

\section{Details on perturbation theory}
\label{APPA}
Here, we explain with some details how one can obtain the series expansions (\ref{EXPANSN}) and (\ref{LARGEZ}), and how the former is related to the DL expansion (\ref{solution LD serie chapitre SSS}). Our starting point is the system 
(\ref{EOMFULL}) rewritten as in (\ref{expansion separe lin quad}). 

Looking for an expansion of the form (\ref{dvt l n m chapitre SSS}), and assuming such an expansion has been found up to order $n-1$ (with $n \geq 1$), the  $n$ th order system is of the form
\be
\label{systeme ordre n expansion loin}
\begin{aligned}
\frac{\lambda'_{n}}{z}+\frac{\lambda_{n}}{z^{2}} + \frac{1}{2}(\lambda_{n}+3\mu_{n}+ z \mu'_{n}) &= f_{t,n}-G_{tt,n},\\
\frac{\nu'_{n}}{z}-\frac{\lambda_{n}}{z^{2}}-\frac{1}{2}(\nu_{n}+2\mu_{n}) &= f_{R,n}-G_{RR,n}, \\
\frac{\lambda_{n}}{z^{2}}-\frac{\nu'_{n}}{2z} &=\frac{1}{z} f_{g,n},
\end{aligned}
\ee
where the functions $ f_{t,n}$, $ f_{R,n}$, $ f_{g,n}$, $G_{tt,n}$ and $G_{RR,n}$ correspond respectively to the terms of order  $\epsi^{n+1}$ of  $ f_{t}$, $ f_{R}$, $f_{g}$, $G_{tt}$ and $G_{RR}$.
It is possible to reorganize the system (\ref{systeme ordre n expansion loin}) to obtain the equivalent system
\ba
\nu'_{n} &=& \frac{2 \lambda_{n}}{z} -2  f_{g,n}, \label{NUNEQ} \\
\lambda_{n} &=& - \frac{z^2 \mu_{n} + z^3 \mu'_{n}}{2+z^2}\nonumber\\
&&-\frac{1}{3(2+z^2)}\left[-2 z^2 f_{t,n}+ 2 z^3 f'_{R,n}+2z^2 G_{tt,n} - 2 z^3 G_{RR,n}'\right.\nonumber\\
&& \left. \qquad-2z(2+z^2) f_{g,n}+ 4 z^3 f_{g,n}'\right],\label{LNEQ}\\
L[\mu_{n}] &=& - 2 z (4+z^2)f_{t,n}+ 2 (2+z^2) f_{t,n}' - (16+6 z^2 + z^4) f_{R,n}' - (4z+2z^3) f_{R,n}''\quad \nonumber\\
&& + 2 z(4+z^2)G_{tt,n} - (4+ 2z^2) G_{tt,n}'     + (16+6 z^2 + z^4) G_{RR,n}' +(4z + 2z^3) G_{RR,n}''\nonumber\\
&&+\frac{(2+z^{2})^{2}(4+z)}{z^{2}} f_{g,n} - \frac{16}{z}f_{g,n}'  - (8 + 4z^2)f_{g,n}''\label{Eq_mu_n1}.
\ea
where $L$ stands for the linear differential operator
\be
L[f]\equiv-3z(4+z^2) f(z) + 6(z^2+4) f'(z) + 3z (z^2+2) f''(z).\nonumber
\ee 
Let's first focus on equation (\ref{Eq_mu_n1}), which allows to find  $\mu_{n}$. This equation can be written in the schematic form
\be\label{Eq_mu_n1_scheme}
L[\mu_{n}]=F[\{\mu_{m < n},\lambda_{m < n},\nu_{m < n}\}],
\ee
where $F$ gathers the nonlinear terms of the equation of order $\epsi^{n+1}$, which only depend on the terms of order $m < n$. 
Note that the above discussion holds both for the large $z$ and the small $z$ expansions, since it is only based on the matching between terms of the same order in the parameter $\epsilon$ on the left and right hand sides of equations (\ref{expansion separe lin quad}). To go further, let us consider the action of the linear operator $L$ on one term of the series
(\ref{EXPANSN}) for $\mu_n$ of the form 
\ba
M_{n,i,k} &\equiv& \left( \frac{e^{-z}}{z^5}\right)^{n+1}   z^{i+2}\left(z^5\log z\right)^{n-k}, \nonumber \\
&=& e^{-(n+1) z} z^{i-3-5k} \left(\log z\right)^{n-k},\nonumber
\ea
such that, with those notations, one is looking for an expansion of $\mu_n$ in the form 
\ba \label{SERIEMU}
\mu_n &=& \epsi^{n+1} \sum_{i=0}^{\infty}\sum_{k=0}^{n} \mu_{n,i,k} M_{n,i,k}. 
\ea
Henceforth, we will denote by $S_{n,m}$ the set of formal series
of the form 
\ba \label{DEF Snm}
z^m \epsi^{n+1} \left( \frac{e^{-z}}{z^5}\right)^{n+1}   \sum_{i=0}^{\infty}\sum_{k=0}^{n} s_{n,i,k} \;  z^{i}\left(z^5\log z\right)^{n-k},\nonumber
\ea
where $s_{n,i,k}$ denotes real coefficients. 
With this notation, we see that the expansion for $\mu_n$ given in Eq. (\ref{EXPANSN}) and (\ref{SERIEMU}) belongs to the set $S_{n,2}$, while the expansions for $\lambda_n$ and $\nu_n$ belong to the set $S_{n,4}$. It is easy to check that those expansions hold for $n=0$, as can be seen from equation (\ref{linsolution}). One finds easily that 
\ba
z L[M_{n,i,k}] &= &p_{n,i,k} M_{n,i,k} \nonumber\\
&&+ q_{n,i,k} M_{n,i+1,k} + r_{n,i,k} M_{n,i+2,k} + s_{n,i,k} M_{n,i+3,k} + t_{n,i,k} M_{n,i+4,k} \nonumber\\
&& +(k-n)\left[a_{n,i,k} M_{n,i+5,k+1} + b_{n,i,k} M_{n,i+6,k+1}+ c_{m,i,k} M_{n,i+7,k+1} + d_{n,i,k}M_{n,i+8,k+1}\right] \nonumber\\
&& +3(k-n)(k-n+1)(2 M_{n,i+10,k+2}+ M_{n,i+12,k+2}) \label{LM}
\ea
where one has 
\ba
p_{n,i,k} &=& 6(i-5k)(i-5k -3)\nonumber \\
a_{n,i,k} &=& 6 (3-2i + 10k)\nonumber
\ea
 and $q_{n,i,k},r_{n,i,k},s_{n,i,k}$ and $t_{n,i,k}$ are quadratic  polynomials of $n,i$ and $k$ with integer coefficients, $b_{n,i,k},c_{n,i,k}$ and $d_{n,i,k}$ are linear  polynomial of $n,i$ and $k$ with integer coefficients. The exact expressions of these polynomials will not be needed here (except for $t_{n,i,n}$, see below).
It is then easy to see, that the action of the linear operator $L$ on any series of $S_{n,2}$ results in a new series element of $S_{n,1}$.
At each order $n$, a necessary condition to find a solution for $\mu_{n}$ in the above form (\ref{SERIEMU}) is then that $F[\{\mu_{m < n},\lambda_{m < n},\nu_{m < n}\}]$ can be written as an element of $S_{n,1}$. Let us show that this is indeed the case. To do so, let us first consider the different functions appearing on the right hand side of Eq. (\ref{expansion separe lin quad}). Those functions have different dependencies on  $\lambda, \nu, \mu$ and $z$ that can be read from equations (\ref{EOMFULL1}-\ref{EOMFULL2}) and (\ref{ft}-\ref{fg}) and are given below.

 \begin{itemize} 
  \item the functions $f_{t,\geqslant2}$ and $f_{R,\geqslant2}$ come from the mass term and, as a consequence, can only depend on $\lambda, \nu, \mu,z\mu'$. This means that once expanded in terms of powers of $\lambda, \nu, \mu$ and $z\mu'$, the functions $f_{t,\geqslant2}$ and $f_{R,\geqslant2}$  are sums of terms of the form
$
\lambda^{n_{1}} \nu^{n_{2}} \mu^{n_{3}}(z\mu')^{n_{4}}
$, with $n_1, n_2$ and $n_3$ positive integers and $n_4$ an integer verifying $0\leq n_4\leq 2$.\\
 
\item the functions $G_{tt,n}$ and $G_{RR,n}$ come from the Einstein's tensor, in the gauge (\ref{lammunu}). Their expression is given by the left hand side of the equations (\ref{EOMFULL1}) and (\ref{EOMFULL2}); one can then easily realize that the functions  $G_{tt,n}$ and $G_{RR,n}$ can be written as an infinite sum of terms of the form
$
\lambda^{n_{1}} \nu^{n_{2}} z^{-2}$, $\lambda^{n_{1}} \nu^{n_{2}}\nu' z^{-1}$ or $\lambda^{n_{1}} \nu^{n_{2}}\lambda' z^{-1}$,  with $n_1$ and $n_2$ positive integers.\\

\item the function $f_{g}$ corresponds to the Bianchi identity (\ref{BIANCHIFULL}), \emph{i.e.} to the derivative of the mass term; it is an infinite sum of terms of the form 
$z^{-1} \lambda^{n_{1}}\nu^{n_{2}}\mu^{n_{3}} (z\mu')^{n_4} (z\nu')^{n_5} (z\lambda')^{n_6} (z^2 \mu'')^{n_7}$ where $n_{i}$ ($i=1...7$) are positive integers and $0\leq n_4 \leq 3$, and  $0\leq n_i\leq 1$ for $5 \leq i\leq 7$.

\end{itemize}

It is then clear that, if one assumes that the series expansions hold at order $\epsi^k$, with $k <n$,  the right hand side of equation (\ref{Eq_mu_n1}) 
can be written as a series of the form 
\ba 
\epsi^{n+1} \sum_{i=i_{min}}^{\infty}\sum_{k=k_{min}}^{n} m_{n,i,k} M_{n,i,k}  = \left(\epsi \; e^z\right)^{n+1} \sum_{i=i_{min}}^{\infty}\sum_{k=k_{min}}^{n} m_{n,i,k}
 z^{i-3-5k} \left(\log z\right)^{n-k}
\ea
It remains to show that  $k_{\min}=0$ and $i_{\min}=-1$ for the above series to be in the $S_{n,1}$ set. It is clear that the first relation holds because the highest power of $\log z$ that can possibly appear on the right hand side of (\ref{Eq_mu_n1}) is necessarily strictly smaller than the power of $\epsi$, i.e., $n+1$, this because it is so in each term of the series $S_{n,m}$.
Let us now compute $i_{min}$. To do so, we need to look at the terms on the right hand side of (\ref{Eq_mu_n1}) which are the most singular power of $z$ when $z$ goes to zero. Keeping in mind that the series $S_{n,2}$ are more singular when $z\rightarrow 0$ than the series $S_{n,4}$, looking at the above described decompositions of the functions $f_{t,\geqslant2}$, $f_{R,\geqslant2}$, $G_{tt,n}$, $G_{RR,n}$, and $f_{g}$, and inspecting Eq.  (\ref{Eq_mu_n1}), one sees that the most singular power of $z$ of the right hand sides of this equation, comes from the terms proportional to 
$z^{-2} f_{g,n}$, $z^{-1}  f'_{g,n}$, and $f''_{g,n}$; and more specifically from products of $\mu^{n_i}$, $(z \mu')^{n_j}$ and $(z^2 \mu'')^{n_l}$ appearing there (with $n_i, n_j$ and $n_l$ positive integers strictly smaller than $n$). In those products, the piece containing the most singular power of z will be given by terms of the form\footnote{The factor of $z^{-3}$ in front of the expression below comes from a factor $z^{-2}$ in front of $f_{g,n}$ and another factor of $z^{-1}$ inside the expression of $f_{g,n}$}
\be 
z^{-3} \prod_{j=1}^{j=J} \left(\epsi^{n_j+1} M_{n_j,0,k_j}\right)^{N_j} \nonumber
\ee
with $N_j$, $n_j$, $k_j$ and $J$ positives integers verifying $k_j \leq n_j < n$, $J\geq 2$ and 
\be \label{EQNn}
\sum_{j=1}^{j=J} (n_j+1)N_j = n+1,
\ee
where the above equality selects the order $\epsi^{n+1}$.
The above expression (\ref{MOSTSING}) can be rewritten as 
\be
(\epsi \; e^z)^{n+1}  z^{-3-\sum_{j=1}^{j=J} (3 + 5 k_j) N_j } \left(\log z\right)^{n+1 -\sum_{j=1}^{j=J} (1 +  k_j) N_j }.\nonumber
\ee
Defining then $k$ and $i$ as 
\ba
k &=& \sum_{j=1}^{j=J} (1 +  k_j) N_j -1\nonumber\\
i &=& -5+2\sum_{j=1}^{j=J} N_j,\nonumber
\ea
the above expression reads now
\ba \label{MOSTSING}
(\epsi \; e^z)^{n+1}  z^{ i-3-5k} \left(\log z\right)^{n-k}.
\ea
With the above definition, $i$ takes its minimal value equal to $-1$ for $\sum_{j=1}^{j=J} N_j =2$, which is achieved for $J=2$ and $N_1=N_2=1$ (for which there is always a solution to equation (\ref{EQNn})). This proves that the right hand side of Eq. (\ref{Eq_mu_n1}) is in the $S_{n,1}$ set, as announced. Moreover, taking now into account the logarithm, it is easy to see what are the most singular terms in the right hand side of Eq. 
(\ref{Eq_mu_n1}) (as z goes to zero). Those are obtained by the terms (\ref{MOSTSING}) with $i=-1$ and $k=n$, which is achieved again under the same conditions as above: $\sum_{j=1}^{j=J} N_j =2$, $J=2$ and $N_1=N_2=1$ (with the additional condition that $\forall j,\; k_j=n_j$). This shows in fact that those most singular terms are simply given by the serie expansion of the Decoupling Limit. Indeed, taking this limit, the only terms which are left in the 
 right hand side of Eq. 
(\ref{Eq_mu_n1}) are those in $f_g$ which are quadratic in $\mu$. The same reasoning also shows that the non linear pieces on the right hand side of Eqs. (\ref{NUNEQ}) and (\ref{LNEQ}) are respectively in the $S_{n,3}$ and $S_{n,4}$ families and have their most singular piece at small $z$ given by the part of $f_{g,n}$ which is quadratic in $\mu$.

This first shows that a (formal) solution of the system (\ref{NUNEQ}-\ref{Eq_mu_n1}) can a priori be searched in the form of the series (\ref{EXPANSN}), but also that, provided such a solution can be found, a subseries of it which corresponds to the most singular terms at small $z$ will be given {\it exactly} by the series expansion obtained in the Decoupling Limit (Eq. (\ref{DLsolinf})).

Let us now show explicitly how to obtain the coefficients $\mu_{n,i,k}$, and solve Eq. (\ref{Eq_mu_n1}) for $\mu_{n}$.
It is first of interest to look at the expression (\ref{LM}) for $k=n$ which reads 
\ba
z L[M_{n,i,n}] &= &6(i-5n)(i-5n -3) M_{n,i,n} \nonumber\\
&&+ q_{n,i,n} M_{n,i+1,n} + r_{n,i,n} M_{n,i+2,n} + s_{n,i,n} M_{n,i+3,n}\nonumber \\
&& +(6n + 3n^2) M_{n,i+4,n} \label{LMN}
\ea
The necessity to incorporate logarithms in the series expansion for $\mu_n$ clearly appears above. Indeed, imagine we look for an expansion without logarithms, i.e. at each order n, we look for an expansion for $\mu_n$, $\nu_n$ and $\lambda_n$ of the form 
\be 
\label{SERIEMUbis}
\begin{aligned}
\mu_n &\sim \epsi^{n+1} \sum_{i=0}^{\infty} m_{n,i} M_{n,i,n}  \\
&= \epsi^{n+1} \sum_{i=0}^{\infty} m_{n,i} e^{-(n+1)z}z^{i-3-5n} \\
\nu_n &\sim \epsi^{n+1} z^2 \sum_{i=0}^{\infty} n_{n,i} M_{n,i,n} \\
\lambda_n &\sim \epsi^{n+1} z^2 \sum_{i=0}^{\infty} l_{n,i} M_{n,i,n} ,
\end{aligned}
\ee
Note that this form is equivalent to the form (\ref{SERIEMU}) for $n=0$, and it is in agreement with the zero order solution given in equation (\ref{linsolution}), and we have proven above that those series must start from $i=0$. Then, at fixed $n$, assuming the above form holds for terms of order $\epsilon^k$ with $k \leq n$, and matching formally the terms proportional to $z^{-1} \times M_{n,i,n}$ on the left hand side and right hand side of equation (\ref{Eq_mu_n1_scheme}), one can try to solve for the values of the coefficients $m_{n,i}$, starting from $i=0$ to increasing values of $i$. This is possible only until one reaches $i=5n$ where the coefficient in front of $M_{n,i,n}$ in (\ref{LMN}) vanishes. This vanishing means that the coefficient in front of the term $z^{-1} \times M_{n,5n,n}$ on the left hand side of (\ref{Eq_mu_n1_scheme}) will not depend on $\mu_{n,5n}$. On the other hand, this coefficient will only depend on the coefficients $\mu_{n,i}$ with $i<5n$ which are already determined by matching the terms $M_{n,i,n}$, with $i<5n$, and one has thus no freedom left to match the term proportional to $M_{n,5n,n}$ on the two sides of (\ref{Eq_mu_n1_scheme}). This  means that it is not possible in general to find for $\mu_n$ a series expansion of the above form (\ref{SERIEMUbis}). Note that the coefficient in front of $M_{n,i,n}$ in (\ref{LMN}) also vanishes for $i=5n+3$, and hence one has the same problem for finding the coefficient $\mu_{n,5n+3}$. Including logarithms however solves this problem and allows to get a formal series solution in the family $S_{n,2}$ as we know show.

Indeed, let us fix $n$ and see how one can compute the coefficients $\mu_{n,i,k}$. Let us first assume that all the coefficients  $\mu_{n,i,k}$, with $k<K$ ($K$ some chosen positive integer) and $i<5k$ are known, and see how to obtain the coefficients  $\mu_{n,i,K}$, with $i<5K$. This is immediate, indeed, the terms proportional to $z^{-1} \times M_{n,i,K}$ (for some $i<5K$) on the left hand side of equation (\ref{Eq_mu_n1_scheme}) depend only on the coefficients $\mu_{n,i,k}$, with $k\leq K$ and $i<5k$, as can be seen from Eq. (\ref{LM}). Hence, by matching formally those terms on the left hand side and right hand side of equation (\ref{Eq_mu_n1_scheme}) one obtains a system of $5K$ equations for the $5K$ unknown coefficients $\mu_{n,i,K}$, with $i<5K$. This system is invertible because, as follows from equation 
(\ref{LM}) it is represented by a triangular matrix with diagonal elements, given by the $p_{n,i,K}$,  which are non vanishing (this because we assumed $i<5K$). Hence one can solve for the $\mu_{n,i,K}$, and by recursion one obtain all coefficients $\mu_{n,i,k}$, with $i<5k$. The coefficients $\mu_{n,5k,k}$ can then be determined as follows: $\mu_{n,0,0}$ is determined uniquely by matching the term proportional to $z^{-1} \times M_{n,5,1}$ on the left hand side and right hand side of equation (\ref{Eq_mu_n1_scheme}). Indeed, $\mu_{n,5,1}$ does not appear in this term, since $p_{n,5,1}$ vanishes, but it is replaced there by $\mu_{n,0,0}$ because $a_{n,0,0}$ does not vanish. By recursion, one can obtain in a similar way all coefficients $\mu_{n,5k,k}$, with $k<n$, indeed, one can check that the coefficients $a_{n,i=5k,k}=18$ and hence do not vanish, which means that at fixed $K$, $\mu_{n,5(K-1),K}$ is uniquely determined by looking at the term $z^{-1} \times M_{n,5K,K}$. The last coefficient $\mu_{n,5n,n}$ is left undetermined and can be chosen at will being related to a zero mode, as we will see below. The coefficients $\mu_{n,i,k}$, with $5k<i<5k+3$ are then determined in a similar manner as those with $i<5k$. One then obtain the $\mu_{n,i=5k+3,k}$ in way similar to the way one obtained the $\mu_{n,5k,k}$ (checking again that $a_{n,i=5k+3,k}=-18$ do not vanish). One is left with one arbitrary choice of $\mu_{n,5n+3,n}$, and the remaining coefficients $\mu_{n,i,k}$, with $i>5k+3$ are obtained inverting an infinite triangular matrix (or a finite one, if one restricts oneself to some maximum value of $i$). The two arbitrary constants to be fixed each correspond to one of the two independent zero modes of the operator $L$. One of this zero modes is $\mu_0$ given in Eq. (\ref{linsolution}), another independent zero mode is given by 
\be
\mu_{0,bis} = \frac{1}{z}\left(1+\frac{1}{z}+\frac{1}{z^2}\right) e^{-z} -\frac{1}{z}\left(1-\frac{1}{z}+\frac{1}{z^2}\right) e^{z}.\nonumber
\ee
One can check that both functions $\mu_{0}$ and $\mu_{0,bis}$ can be expanded as in (\ref{SERIEMU}) with all coefficients $\mu_{n,i,k}$ vanishing but those with $k=n,$ and $i \geq 5n$ for $\mu_0$ and those with   $k=n,$ and $i \geq 5n +3 $ for $\mu_{0,bis}$. Note, as discussed in the main text, that the inclusions of such zero modes in the series expansions might be necessary to enforce the hierarchy $\mu_{n} \ll \mu_{n-1}$.

Having shown how one can obtain $\mu_n$ in the announced form, it is easy to obtain the corresponding expansions for $\lambda_n$ and $\nu_n$ from equations (\ref{NUNEQ}) and (\ref{LNEQ}). There, there is no free constant left to be chosen.

Let us now turn to the asymptotic expansion (\ref{LARGEZ}) for large $z$.
Following a similar reasoning as previously, it is first easy to see that, provided such an expansion holds up to order $\epsi^{n}$, then the right hand side of equation (\ref{Eq_mu_n1_scheme}) will be a series of the form 
\be
\epsi^{n+1} e^{-(n+1)z}\sum_{i=-\infty}^{i=3} m_{n,i} z^i.\nonumber
\ee
But one also sees from equation (\ref{LMN}) that this form is also the one resulting from the action of the linear operator $L$ on the series  for $\mu_{n}$ given  in (\ref{LARGEZ}). The expression (\ref{LMN}) then shows that one can solve uniquely for all the coefficients $\mu_{n,i}$ starting from the largest values of $i$, and no free constants appear in this case because the coefficient $t_{n,i,n} = 6n + 3n^2$ does not vanish. One can then easily obtain the expressions for $\nu_n$ and $\lambda_n$ in the form  (\ref{LARGEZ}) with no extra free constant.

\section{Series inside the source}\label{series inside the source}
The coefficients of the series expansions (\ref{expansion in x}) can be found solving order by order in $x^{2}$; the first orders read: 
\be
\begin{aligned}
&\mu(x)=\mu_{0}^{(x)}+
\frac{1 }{20 \left(2 e^{\nu_{0}^{(x)}}+e^{\mu_{0}^{(x)}}-3 e^{\nu_{0}^{(x)}+\mu_{0}^{(x)}}\right)}\\
&
\qquad\times\Bigg[ \left(3+4 e^{-\nu_{0}^{(x)}}-39 e^{\nu_{0}^{(x)}}-2 e^{\nu_{0}^{(x)}-2 \mu_{0}^{(x)}}+18 e^{\nu_{0}^{(x)}-\mu_{0}^{(x)}}\right.\\
&
\qquad\qquad\left.-2 e^{-\mu_{0}^{(x)}}-e^{\mu_{0}^{(x)}}-5 e^{\mu_{0}^{(x)}-\nu_{0}^{(x)}}+24 e^{\nu_{0}^{(x)}+\mu_{0}^{(x)}}\right) m^2\\
&
\qquad\qquad+2 \left(3 e^{\nu_{0}^{(x)}+\mu_{0}^{(x)}} \left(P_{0}^{(x)}-\frac{5}{x_{\odot}^3}\right)-2 e^{\nu_{0}^{(x)}} \left(P_{0}^{(x)}-\frac{3}{x_{\odot}^3}\right)+e^{\mu_{0}^{(x)}} \left(P_{0}^{(x)}+\frac{7}{x_{\odot}^3}\right)
\right)\Bigg]\;x^{2}+\mathcal{O}(x^{4})\nonumber
\end{aligned}
\ee

\be
\begin{aligned}
\lambda(x)=\frac{1}{4} \left(\frac{4}{x_{\odot}^3}-e^{-\nu_{0}^{(x)}-2 \mu_{0}^{(x)}} \left(-1+e^{\mu_{0}^{(x)}}\right) \left(-e^{\nu_{0}^{(x)}}+e^{\mu_{0}^{(x)}}+2 e^{\nu_{0}^{(x)}+\mu_{0}^{(x)}}\right) m^2\right)x^{2}+\mathcal{O}(x^{4})\nonumber
\end{aligned}
\ee

\be
\begin{aligned}
\nu(x)=& \nu_{0}^{(x)}+\\
&+\frac{1}{4} e^{-\nu_{0}^{(x)}-2 \mu_{0}^{(x)}} \left(-e^{\nu_{0}^{(x)}} m^2+e^{\mu_{0}^{(x)}} m^2-2 e^{2 \mu_{0}^{(x)}} m^2+2 e^{\nu_{0}^{(x)}+2 \mu_{0}^{(x)}} \left(m^2+P_{0}^{(x)}+\frac{1}{x_{\odot}^3}\right)\right)x^{2}+\mathcal{O}(x^{4}) \nonumber
\end{aligned}
\ee

\be
\begin{aligned}
P(x)&=P_{0}^{(x)}-\frac{1}{8} \left(P_{0}^{(x)}+\frac{3}{x_{\odot}^3}\right)e^{-\nu_{0}^{(x)}-2 \mu_{0}^{(x)}}\\
&\quad\quad\times  \left(-e^{\nu_{0}^{(x)}} m^2+e^{\mu_{0}^{(x)}} m^2-2 e^{2 \mu_{0}^{(x)}} m^2+2 e^{\nu_{0}^{(x)}+2 \mu_{0}^{(x)}} \left(m^2+P_{0}^{(x)}+\frac{1}{x_{\odot}^3}\right)\right)x^{2} +\mathcal{O}(x^{4}) \nonumber
\end{aligned}
\ee

\section{Infinitely many solutions at infinity: a series expansion approach}\label{Infinitely many solutions at infinity: a series expansion approach}
In this Appendix, we would like to complete the argument on the existence, at infinity,  of infinitely many solutions of the system (\ref{EOMFULL}) having the same asymptotic behaviour  given by the linear solution  (\ref{linsolution}). Our starting point is section \ref{NUMINF}, where we assumed that there exists a solution
$\bar\sigma\equiv\{\bar\lambda,\bar\nu,\bar\mu\}$, such that 
\be
\sigma\to\sigma_0  \quad \text{for} \quad z\to\infty. \nonumber
\ee
where $\sigma_0\equiv\{\lambda_0,\nu_0,\mu_0\}$, we looked for a solution  
$\sigma$ as small perturbations of the known solution  $\bar{\sigma}$:
\be
\label{delta1}
\sigma=\bar\sigma+\delta\sigma,\nonumber
\ee
where $\delta\sigma\equiv\{\delta\lambda,\delta\nu,\delta\mu\}$. We gave the leading behaviour of these solutions at Eq. (\ref{finalsolz}). In the following, we would like to 
 go beyond this leading behaviour analysis and find a consistent way of looking for the solution $\delta \sigma$ through a series expansion. 
 
 \subsection{Equations for $\delta \sigma$}
 The first step is to write down the equations for $\delta \sigma$, that would generalize the leading order equations (\ref{lindelta1}).
Schematically, taking into account (\ref{assumedelta}), the system of equations (\ref{EOMFULL}) can be linearized 
around the solution $\bar\sigma$,
\be
\label{lindelta}
\begin{aligned}
\frac{\partial F_t}{\partial\sigma}\Big|_{\bar\sigma}\delta\sigma
+\frac{\partial F_t}{\partial\sigma'}\Big|_{\bar\sigma}\delta\sigma'&=0,\\
\frac{\partial F_R}{\partial\sigma}\Big|_{\bar\sigma}\delta\sigma
+\frac{\partial F_R}{\partial\sigma'}\Big|_{\bar\sigma}\delta\sigma'&=0,\\
\frac{\partial f_g}{\partial\sigma}\Big|_{\bar\sigma}\delta\sigma
+\frac{\partial f_g}{\partial\sigma'}\Big|_{\bar\sigma}\delta\sigma'
+\frac{\partial f_g}{\partial\mu''}\Big|_{\bar\sigma}\delta\mu''&=0.
\end{aligned}
\ee
It is important to note that in the linearized Bianchi identity, the last equation of (\ref{lindelta}), 
we do not assume that the terms containing $\delta\mu''$ are negligible, so we keep them.
We are interested in asymptotic solutions at infinity, 
thus we can substitute $\sigma_0$ instead of $\bar\sigma$ in (\ref{lindelta}), since
$\bar\sigma\to\sigma_0$ at $z\to\infty$.

Further, since we are considering a range of distances where the solution is deep in the linear regime, we can safely keep only the lowest order in powers of $\sigma_0$, while making an expansion of coefficients of equations (\ref{lindelta}). This leads to the equations, 
\be
\label{eq delta sigma}
\begin{aligned}
\frac{\delta\lambda'}{z}+\frac{\delta\lambda}{z^2}+\frac12\left(\delta\lambda+3\delta\mu+z \delta\mu'\right)&=
be^{-z}\left[\left(\frac{1}{3z^{3}}-\frac{1}{z^{2}}-\frac{2}{3z}\right)\delta\lambda +\left(\frac{2}{z^{2}}+\frac{2}{3z}\right)\delta\lambda' \right.\\
&\quad\left.+\frac{1}{3z}\delta\nu+\frac{1}{z}\delta\mu+\left(-\frac{1}{2z^{2}}-\frac{1}{2z}+\frac{1}{3}+\frac{z}{6}\right)\delta\mu' \right],\\
\frac{\delta\nu'}{z}-\frac{\delta\lambda}{z^2}-\frac12\left(\delta\nu+2 \delta\mu\right)&=
be^{-z}\left[\left(\frac{1}{z^{3}}+\frac{1}{z^{2}}\right)\delta\lambda \right.\\
&\quad\left.+\left(\frac{2}{3z^{3}} +\frac{2}{3z^{2}}+\frac{4}{3z}+\frac{1}{3}\right)\delta\nu\right.\\
&\quad\left.+\left(-\frac{1}{3z}+\frac{2}{3}\right)\delta\mu+\left(-\frac{1}{3z^{2}}-\frac{1}{3z}\right)\delta\mu' \right],\\
\frac{\delta\nu'}{2z}-\frac{\delta\lambda}{z^{2}}&=
\frac{be^{-z}}{z^{2}}\left[\left(\frac{1}{z^{3}}+\frac{1}{z^{2}}-\frac{2}{3z}+\frac{1}{3}\right)\delta\lambda +\left(\frac{1}{3z^{2}}+\frac{1}{3z}\right)\delta\lambda' \right.\\
&\quad\left.+\left(\frac{2}{3z}+\frac{2}{3}\right)\delta\nu+\left(-\frac{1}{z^{2}}-\frac{1}{z}-\frac{2}{3}\right)\delta\nu'\right.\\
&\quad\left.+\left(-\frac{2}{3z}-\frac{2}{3}\right)\delta\mu+\left(\frac{1}{3z^{2}}+\frac{1}{3z}-\frac{1}{3}\right)\delta\mu' \right.\\
&\quad\left.+\left(\frac{1}{3z}+\frac{1}{3}\right)\delta\mu''\right].
\end{aligned}
\ee
One can of course check that keeping only the leading behaviour in the limit $z\to \infty$ leads to the equations (\ref{lindelta1}).

\subsection{Ansatz for $\delta \sigma$}
The next step is to adopt an appropriate  ansatz, which can be chosen of the form
\be
\label{ansatze G}
\begin{aligned}
\delta\mu&=F_{\mu}(z)\exp\left(\frac{z}{4}-f(z)\;e^{z/2}\right),\\
\delta\lambda&=-F_{\lambda}(z)\frac{z^2}{2}\exp\left(\frac{z}{4}-f(z)\;e^{z/2}\right),\\
\delta\nu&=-F_{\nu}(z)\frac{\sqrt{b}}{3}\sqrt{1+\frac{1}{z}}\exp\left(-\frac{z}{4}-f(z)\;e^{z/2}\right),
\end{aligned}
\ee
where the function $f$ is assumed to satisfy the equation
\be
f'(z)=\frac{3 z^{3/2}}{2\sqrt{b(1+z)}}-\frac{f(z)}{2}\label{eq f}.
\ee
The form of this ansatz will be justified later. The formal solution of Eq. (\ref{eq f}) is given by
\be
f(z)=e^{-z/2}\int_{z_{0}}^{z}\frac{3\; t^{3/2}}{2\sqrt{b(1+t)}}e^{t/2} \;dt. \nonumber
\ee
The exact choice of the lower integral bound $z_{0}$ does not matter so much, since it will just translate into an overall constant factor in Eqs. (\ref{ansatze G}). An asymptotic expansion of the solution can be easily found to be
\be
f(z)=\frac{1}{\sqrt{b}}\left(3 z-\frac{15}{2}+\frac{9}{8\;z}+...\right).\label{series expansion for f}
\ee
Note that the leading behaviour of the ansatz (\ref{ansatze G}) matches the asymptotic behaviour (\ref{finalsolz}).

\subsection{The equations for the $F_{i}$'s}
Once the  ansatz (\ref{ansatze G}) has been chosen, we can compute the equations for the $F_{i}$'s. They read\footnote{Note that we have multiplied the $tt$ equation by a factor $\sqrt{b}e^{-z/2}$ in order to have consistent form among the three equations.}
\be
\label{equations Fis}
\begin{aligned}
E_{t}&\equiv E_{t}^{(0)}+\left(\sqrt{b} e^{-z/2}\right)E_{t}^{(1)}+\left(be^{-z}\right)E_{t}^{(2)}+\left(b^{3/2} e^{-3 z/2} \right)E_{t}^{(3)}+\left(b^2 e^{-2 z}\right)E_{t}^{(4)}=0,\\
E_{R}&\equiv E_{R}^{(0)}+\left(\sqrt{b} e^{-z/2}\right)E_{R}^{(1)}+\left(be^{-z}\right)E_{R}^{(2)}+\left(b^{3/2} e^{-3 z/2} \right)E_{R}^{(3)}=0,\\
E_{B}&\equiv E_{B}^{(0)}+\left(\sqrt{b} e^{-z/2}\right)E_{B}^{(1)}+\left(be^{-z}\right)E_{B}^{(2)}+\left(b^{3/2} e^{-3 z/2} \right)E_{B}^{(3)}=0,
\end{aligned}
\ee
with
\be
\label{def Es}
\begin{aligned}
E_{t}^{(0)}\left[F_{i}\right]&\equiv \frac{3 z^{5/2}}{4 \sqrt{z+1}}\left[F_{\lambda}(z)-F_{\mu}(z)\right], \\
E_{t}^{(1)}\left[F_{i}\right]&\equiv -\frac{1}{8} \left[2 \left(z^2+ z+12\right) F_{\lambda}(z)+4 zF_{\lambda}'(z) -(12 +z)F_{\mu}(z)-4z F_{\mu}'(z) \right], \\
E_{t}^{(2)}\left[F_{i}\right]&\equiv \frac{1}{4 \sqrt{z} \sqrt{z+1}}\left[-2 z^2 (z+3) F_{\lambda}(z)+\left(z^3+2 z^2-3 z-3\right)F_{\mu}(z)\right],\\
E_{t}^{(3)}\left[F_{i}\right]&\equiv -\frac{1}{24 z^2}\left[(6z^3-10  z^2-52 z)F_{\lambda}(z)-( 8 z^3+24 z^2)F_{\lambda}'(z)\right.\\
&
\left.+(z^3+2  z^2+21 z-3)F_{\mu}(z) +(4 z^3+8 z^2-12 z-12 )F_{\mu}'(z)\right],\\
E_{t}^{(4)}\left[F_{i}\right]&\equiv \frac{1}{9 z^2}\sqrt{z^2+z}\; F_{\nu}(z),\\
\\
E_{R}^{(0)}\left[F_{i}\right]&\equiv \frac{1}{2}\left[F_{\lambda}(z)+F_{\nu}(z)-2 F_{\mu}(z)\right] ,\\
E_{R}^{(1)}\left[F_{i}\right]&\equiv \frac{1}{12 z^{5/2} \sqrt{z+1}} \left[\left(2z^{3}+3z^2+ z+2\right) F_{\nu}(z) \right.\\
&\left.-4( z^{2}+z)F_{\nu}'(z) -6(z^{3} +z^{2})F_{\mu}(z)\right], \\
E_{R}^{(2)}\left[F_{i}\right]&\equiv \frac{1}{12z^{2}}\left[6 z (1+z) F_{\lambda}(z)+(1+5 z-8 z^2)F_{\mu}(z)+4 (1+z)F_{\mu}'(z)\right],\\
E_{R}^{(3)}\left[F_{i}\right]&\equiv \frac{1}{9 z^{7/2} \sqrt{z+1}}\left[(z^4+5 z^3+6 z^2+4 z+2)F_{\nu}(z)\right],\\
\\
E_{B}^{(0)}\left[F_{i}\right]&\equiv \frac{1}{4}\left[2F_{\lambda}(z)+F_{\nu}(z)-3 F_{\mu}(z)\right], \\
E_{B}^{(1)}\left[F_{i}\right]&\equiv - \frac{1}{24 \sqrt{z^5 (z+1)}}\left[-6 (z^{2}+z^{2})F_{\lambda}(z)+\left(z^2+z+2\right) F_{\nu}(z)-4( z+z^{2})F_{\nu}'(z)\right.\\
&\left. +(36z+30)F_{\mu}(z) +24(z+ z^2 )F_{\mu}'(z)\right], \\
E_{B}^{(2)}\left[F_{i}\right]&\equiv \frac{1}{48 z^{4}}\left[2 z\left(4 z^3-7 z^2+21 z+20\right) F_{\lambda}(z) +(8 z^3+8z^2)F_{\lambda}'(z) \right.\\
&\left. +8 z\left(2 z^2+3 z+3\right) F_{\nu}(z)+(35  z^2+27  z-4) F_{\mu}(z)\right.\\
&\left. +(8  z^2-24 z-16 ) F_{\mu}'(z)-16 ( z^2-16z) F_{\mu}''(z) \right],\\
E_{B}^{(3)}\left[F_{i}\right]&\equiv \frac{1}{36 z^5 \sqrt{z (z+1)}}\left[\left(\!10 z^4+21 z^3+18 z^2+9 z+6\right)F_{\nu}(z)\right.\\
&\left. -4 z \left(2 z^3+5 z^2+6 z+3\right)\!F_{\nu}'(z)\right].
\end{aligned}
\ee
Note that we have used Eq. (\ref{eq f}) to eliminate $f'$ and $f''$; surprisingly enough, the equations (\ref{def Es}) do not contain $f$ either.

\subsection{Expansion in powers of $\sqrt{b}\;e^{-z/2}$}
The next step is to solve for the functions $F_{i}$. To do so, let's assume they are of the form
\be
F_i(z)=\sum_{n=0}^{\infty}F_{i}^{(n)}(z)\;\left(\sqrt{b}\;e^{-z/2}\right)^{n}\label{series form for Fi}
\ee
with $F_{i}^{(n)}(z)$ at most polynomial, i.e. 
\be
F_{i}^{(n)}(z)=\mathcal{O}\left(z^{p_{i}^{(n)}}\right),\label{polynomial condition on Fi}
\ee
and solve the equations (\ref{eq delta sigma}) order by order in  $\sqrt{b}e^{-z/2}$.

\paragraph{$0$-th order}
A priori, out of the three equations (\ref{equations Fis}), one can get the three $0$-th order equations for $F_{i}^{(0)}$
\be
E_{k}^{(0)}\left[F_{i}^{(0)}\right]=0, \nonumber
\ee
where $k=\{t,R,B\}$ and $i=\{\lambda,\nu,\mu\}$.
However, only two of them are independent. To see this more algebraically, one can write the $0$-th order equations in a matrix form
\be
\mathbf{M.F}^{(0)}=0, \nonumber
\ee
where
\be
\mathbf{M}=\left(\begin{array}{ccc}\frac{3 z^{5/2}}{4 \sqrt{z+1}} & 0 & -\frac{3 z^{5/2}}{4 \sqrt{z+1}} \\\frac{1}{2} & \frac{1}{2} & -1 \\\frac{1}{2} & \frac{1}{4} & -\frac{3}{4}\end{array}\right)\quad\text{and}\quad \mathbf{F}^{(0)}=\left(\begin{array}{c}F_{\lambda}^{(0)}\\F_{\nu}^{(0)}\\F_{\mu}^{(0)}\end{array}\right). \nonumber
\ee
The vector $\mathbf{V}_{Ker}=\left(\begin{array}{c}\frac{2}{z}\\\frac{3 z^{3/2}}{ \sqrt{z+1}} \\-\frac{6 z^{3/2}}{ \sqrt{z+1}}\end{array}\right)$ belongs to the kernel of $\mathbf{M}^{\top}$, while the vectors
\be
\mathbf{V}_{\lambda}=\left(\begin{array}{c}\frac{4 \sqrt{z+1}}{3 z^{5/2}}\\0\\0\end{array}\right)\quad,\quad\mathbf{V}_{\nu}=\left(\begin{array}{c}-\frac{4 \sqrt{z+1}}{3 z^{5/2}}\\2\\0\end{array}\right) \nonumber
\ee
can be used to get the equations
\be
\label{replace order 0}
\begin{aligned}
\mathbf{V}^{\top}_{\lambda}.\mathbf{M.F}^{(0)}&= F^{(0)}_{\lambda}(z)-F^{(0)}_{\mu}(z)=0,\\
\mathbf{V}^{\top}_{\nu}.\mathbf{M.F}^{(0)}&= F^{(0)}_{\nu}(z)-F^{(0)}_{\mu}(z)=0.
\end{aligned}
\ee
The fact that $\mathbf{M}$ is of rank 2 is a direct consequence of the  ansatz (\ref{ansatze G}), and in particular of the differential equation (\ref{eq f}) satisfied by $f$. More precisely, if one does not assume anything on the function $f$, and then computes the $0$-th order equations out of the ansatz (\ref{ansatze G}), they are still of the homogeneous form $\mathbf{N.F}^{(0)}=0$, but with the matrix $\mathbf{N}$ given by
\be
\mathbf{N}=\left(\begin{array}{ccc}\frac{\sqrt{b}}{4}z\left(f(z)+2f'(z)\right) & 0 &-\frac{\sqrt{b}}{4}z\left(f(z)+2f'(z)\right)  \\ \frac{1}{2} & \frac{\sqrt{b} \sqrt{z (z+1)} \left(f(z)+2 f'(z)\right)}{6 z^2} & -1 \\ \frac{1}{2} & \frac{\sqrt{b} \sqrt{z (z+1)} \left(f(z)+2 f'(z)\right)}{12 z^2} & -\frac{b(z+1) \left(f(z)+2 f'(z)\right)^2}{12 z^3}\end{array}\right). \nonumber
\ee
In order to have $\mathbf{N}$ of rank 2 (otherwise $\mathbf{F}^{(0)}$ is trivially $0$), it is necessary to impose that its determinant is $0$. If in addition we require $f$ to be positive, the condition $\det \mathbf{N}=0$ translates into the differential equation (\ref{eq f}).

To conclude, we have found that the  $0$-th order equations are not enough to fix the $F_{i}^{(0)}$'s; however, they allow to understand the origin of the condition (\ref{eq f}) and enforce $F^{(0)}_{\lambda}(z)=F^{(0)}_{\nu}(z)=F^{(0)}_{\mu}(z)$. We now have to go to next order to solve for $F^{(0)}_{\mu}(z)$.

\paragraph{First order} The first order equations read
\be
E_{k}^{(0)}[F_{i}^{(1)}]+E_{k}^{(1)}[F_{i}^{(0)}]=0.\label{eq Fi order 1}
\ee
Let's first project this set of equations onto the Kernel vector $\mathbf{V}_{Ker}$, since we know this procedure will cancel the $E_{k}^{(0)}[F_{i}^{(1)}]$ contribution:
\be
\mathbf{V}_{Ker}^{\top}\;.\;\mathbf{E}^{(1)}[F_{i}^{(0)}]=\frac{3 \sqrt{b}}{2 z (z+1)} \left[(6 z+5) F^{(0)}_{\mu}(z)+4 z (z+1) F_{\mu}^{(0)'}(z)\right]=0 \label{eq Fmu0}
\ee
where we have used the relations (\ref{replace order 0}) to replace $F_{\lambda}, F_{\nu}$ by $F_{\mu}$. The equation (\ref{eq Fmu0}) can be straightforwardly solved, leading to
\be
F^{(0)}_{\lambda}(z)=F^{(0)}_{\nu}(z)=F^{(0)}_{\mu}(z)=\frac{A}{z^{5/4} (1+z)^{1/4}}, \label{solution F0}
\ee
with $A$ being an arbitrary integration constant. A few remarks can be made at this point. First, the arbitrariness of the constant $A$ explicitly shows that out of a solution $\bar\sigma$ of the system of equations (\ref{EOMFULL}), it is possible to find a family of infinite number of new solutions, with the same asymptotic behaviour at infinity. Second, we note that $F_{i}^{(0)}=\mathcal{O}\left(z^{-3/2}\right)$, in agreement with the condition (\ref{polynomial condition on Fi}).

Let's turn now to the two other first order equations. Projecting Eq. (\ref{eq Fi order 1}) onto $\mathbf{V}_{\lambda}$ and $\mathbf{V}_{\nu}$ leads to
\ba
&&F^{(1)}_{\lambda}(z)-F^{(1)}_{\mu}(z)-\frac{1}{3} \sqrt{1+\frac{1}{z}}\;F_{\mu}^{(0)}(z)=0,\nonumber\\
&&F^{(1)}_{\nu}(z)-F^{(1)}_{\mu}(z)+\frac{-2 z^3-z^2+z+2}{6 \sqrt{z^5 (z+1)}}\;F_{\mu}^{(0)}(z)-\frac{4 z^2+4 z}{6 \sqrt{z^5 (z+1)}} F_{\mu}^{(0)'}(z)=0.\nonumber
\ea
These equations allow to replace $F^{(1)}_{\lambda}$ and $F^{(1)}_{\nu}$ by $F^{(1)}_{\mu}$ in the equations of order 2. These order 2 equations will lead to an equation for $F^{(1)}_{\mu}$, and two other equations relating $F^{(2)}_{\lambda}$ and $F^{(2)}_{\nu}$ to $F^{(2)}_{\mu}$, and so on and so forth.

\paragraph{$n$-th order} The $n$-th order equations read
\be
E_{k}^{(0)}[F_{i}^{(n)}]+E_{k}^{(1)}[F_{i}^{(n-1)}]+E_{k}^{(2)}[F_{i}^{(n-2)}]+E_{k}^{(3)}[F_{i}^{(n-3)}]+E_{k}^{(4)}[F_{i}^{(n-4)}]=0.\nonumber
\ee
By iteration, it is possible to prove that
\be
F_{\lambda}^{(n)}(z)\sim F^{(n)}_{\nu}(z)\sim F^{(n)}_{\mu}(z)=\mathcal{O}\left(z^{-3/2+n}\right).\label{asymptotic Fi}
\ee
We can sketch briefly how the demonstration goes: after having solved the equation of order $n$, the functions $F_{i}^{(n-1)}$ are assumed to be known, and $F_{\lambda}^{(n)},F^{(n)}_{\nu}$ are can be expressed as functions of $\{F^{(n)}_{\mu},F_{i}^{(n-1)}\}$. Then, projecting the equations of order $n+1$ onto $\mathbf{V}_{Ker}$ allows to find $F_{\mu}^{(n)}$, leading to the scaling (\ref{asymptotic Fi}). Projecting onto $\mathbf{V}_{\lambda}$ and $\mathbf{V}_{\nu}$ relates $F^{(n+1)}_{\lambda}$ and $F^{(n+1)}_{\nu}$ to $F^{(n+1)}_{\mu}$, closing the iteration process.

To summarize the findings of this section, we can say that we have found a class of solutions for the system of equations (\ref{eq delta sigma}). These solutions are of the form (\ref{ansatze G}), where the functions $F_{i}$ can be expanded into the series of Eq. (\ref{series form for Fi}). There are an infinite number of such solutions. This shows that out of a solution $\bar\sigma$ of the system of equations (\ref{EOMFULL}), it is possible to find an infinite family of new solutions, with the same asymptotic behaviour at infinity. This freedom is crucial to match the vacuum solution with the one with source.

\subsection{Numerical validation}
\begin{figure}[h]
\includegraphics[width=0.5\linewidth]{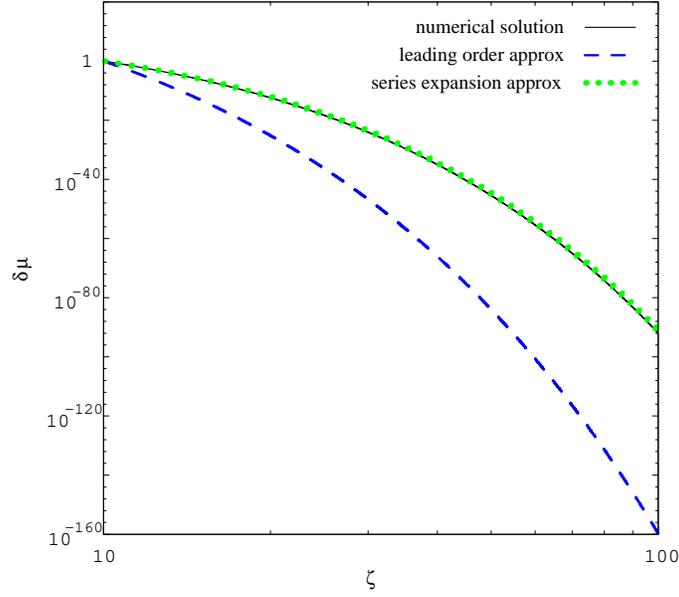} 
\caption{Plot of the numerical solution for $\delta\mu(\zeta)$ for the choice of parameters $b=0.1$ and $\delta\mu(\zeta=10)=1$. We also plotted the leading behaviour $\delta\mu(\zeta)=\exp\left(-3\sqrt{\frac{\zeta}{b}}\log\zeta\right)$ of Eq. (\ref{finalsol}), and the first series expansion correction obtained from Eq. (\ref{series expansion for f}) and Eq. (\ref{solution F0}) . In terms of the $\zeta$ variable, this first order series expansion expression reads $\delta\mu(\zeta)=\frac{1}{(\log\zeta)^{5/4} (1+\log\zeta)^{1/4}}\exp\left(\frac{\log\zeta}{4}-\frac{1}{\sqrt{b}}\left(3 \log\zeta-\frac{15}{2}+\frac{9}{8\;\log\zeta}\right)\;\sqrt{\zeta}\right)$. One can see that while the leading behaviour already encodes the main behaviour (exponentially decreasing $\delta\mu(\zeta)$), the first correction coming from the series expansion improves greatly the agreement with the numerical solution, confirming our analytical analysis of the system (\ref{eq delta sigma}).}
\label{Fig delta sigma}
\end{figure}

To conclude this study of  the infinitely many solutions satisfying the same boundary conditions at infinity,  we checked that our solution  (\ref{ansatze G}) fits well the numerical solution of the system of equations (\ref{eq delta sigma}). An example is shown in Fig. \ref{Fig delta sigma}.

\end{document}